\documentclass[aps,prc,preprint,tightenlines,groupedaddress,showpacs,endfloats]{revtex4} 
\usepackage{graphicx}
\usepackage{longtable}
\usepackage{dcolumn}
\usepackage{amssymb}
\usepackage[bookmarksnumbered,bookmarksopen]{hyperref}

\newcommand{\isotope}[3]{{}_{#3}^{#2}\text{{#1}}}
\newcommand{\eu}{\isotope{Eu}{152m,g}{}}
\newcommand{\eug}{\isotope{Eu}{152g}{}}
\newcommand{\sm}{\isotope{Sm}{152}{}}
\newcommand{\eum}{\isotope{Eu}{152m}{}}
\newcommand{\euz}{\isotope{Eu}{154}{}}
\newcommand{\gd}{\isotope{Gd}{152}{}}

\newcommand{\git}{School of Physics, Georgia Institute of Technology, Atlanta,
Georgia 30332-0430, USA}
\newcommand{\lbnl}{Nuclear Science Division,  Lawrence Berkeley National
Laboratory, Berkeley, California 94720, USA}
\newcommand{\osu}{Department of Physics, Oregon State University,
Corvallis,Oregon 97331-6507, USA}
\newcommand{\lsu}{Department of Physics and Astronomy, Louisiana State University,
Baton Rouge, Louisiana  70803, USA}

\begin{document}

\title{$N=90$ region:  The decays of $\eu$ to $\sm$.}

\author{W.~D.~Kulp}
\author{J.~L.~Wood}
\affiliation{\git}
\author{J.~M.~Allmond}
\author{J.~Eimer}
\author{D.~Furse}
\affiliation{\git}
\author{K.~S.~Krane}
\affiliation{\osu}
\author{R.-M.~Larimer}
\thanks{Deceased}
\affiliation{\lbnl}
\author{J.~Loats}
\altaffiliation[Present address:  ]{Department of Physics and Engineering,
Fort Lewis College, Durango, Colorado 81301, USA}
\affiliation{\osu}
\author{E.~B.~Norman}
\altaffiliation[Present address:  ]{L-414, Lawrence Livermore National
Laboratory, P.O. Box 808, Livermore, CA 94551, USA}
\affiliation{\lbnl}
\author{A.~Piechaczek}
\affiliation{\lsu}
\author{P.~Schmelzenbach}
\altaffiliation[Present address:  ]{Department of Physics and Engineering, 
Point Loma Nazarene University, 3900 Lomaland Drive, San Diego, CA 92106-2899, USA}
\affiliation{\osu}
\author{C.~J.~Stapels}
\altaffiliation[Present address:  ]{Radiation Monitoring Devices, 44 Hunt St., Watertown, Massachusetts  02472, USA}
\affiliation{\osu}

\date{\today}

\begin{abstract}
The decays of $\eu \rightarrow \sm$ have been studied by $\gamma$-ray spectroscopy using the $8\pi$ Spectrometer, an array of 20 Compton-suppressed Ge detectors.  Very weak $\gamma$-decay branches in $\sm$ were investigated through $\gamma - \gamma$ coincidence spectroscopy.  All possible $E2$ transitions between states below 1550 keV with transition energies $> 130$ keV are observed, including the previously unobserved $2^+_3 \rightarrow 0^+_2$ 401 keV transition.  The results, combined with existing lifetime data, provide a number of new or revised $E2$ transition strengths which are critical for clarifying the collective structure of $\sm$ and the $N=90$ isotones.
\end{abstract}

\pacs{21.10.Re, 23.20.Lv, 27.70.+q}

\maketitle


\section{INTRODUCTION}

\paragraph*{}
  The $N=90$ isotones in the vicinity of $Z=64$ are located at the center of a region of rapid change in nuclear shape, and consequently these nuclei demonstrate a rapid change in nuclear collectivity.  
  The samarium isotopes ($Z=62$) particularly have been the subject of many experimental and theoretical studies, to try and understand this change.

\paragraph*{}
  Renewed focus has been brought to this region recently, initiated with studies of very weak $\gamma$-decay branches in $\sm$ \cite{Casten1998}, which has led to the proposition \cite{Iachello1998, Zamfir1999a} that this nucleus has been incorrectly interpreted.  
  The issues are whether or not this nucleus exhibits coexisting spherical and deformed phases \cite{Iachello1998, Zamfir1999a} and whether or not it lies almost exactly at the critical point \cite{Iachello1998} between these phases.  
  Additional data from subsequent studies were presented \cite{Klug2000, Zamfir2002b} to support these interpretations.
  However, there has been controversy \cite{Burke2002a, Clark2003a, Clark2003b, Casten2003} regarding these ideas.

\paragraph*{}
  We have initiated a program of detailed spectroscopy to provide additional data which might clarify the underlying nuclear structure in this region.  
  In $\sm$ a number of critical collective transitions involve low-energy decay branches in competition with high-energy branches which are consequently very weak and difficult to accurately quantify.  
  Indeed, the inception of the new focus on this region involved the non-observation \cite{Casten1998, Zamfir1999a} of the ($J^\pi_i$ ($E_x$ keV)) $2^+_3 (1086) \rightarrow 0^+_2 (685)$ $E2$ transition in $\sm$.

\paragraph*{}
  The $\beta^+/$EC decays of $\eug$ (13.6~year) and $\eum$ (9.3~h) are especially suited to the study of weak $\gamma$-decay branches in $\sm$ because the parent spin-parities of $3^-$ and $0^-$ and the $Q_{\mathrm{EC}}$ values of 1874 and 1920~keV, respectively, confine the $\gamma$-ray spectra to transitions between the low-spin, low-energy collective states that are of interest.  
  This minimizes the problem of intense Compton backgrounds from high-energy $\gamma$ rays, which obscure weak low-energy $\gamma$ rays.
  Thus, we have undertaken very detailed studies of these decays.

\section{\label{procedure}EXPERIMENTAL PROCEDURE}

\paragraph*{}
  The $\eu$ sources were produced by neutron capture on $\sim$97$\%$ enriched $^{151}$Eu oxide in the Oregon State University reactor. 
  Different irradiation and ``cooling'' times were used for the 9.3~h and 13.6~year activities.
  Sources were counted as solutions (in nitric acid) contained in vials of diameter 1~cm $\times$ length 2~cm and, in the case of the 9.3~h activity, were replenished.
Nominal source activities were 50~$\mu$Ci for $\eug$ and 10~$\mu$Ci for $\eum$.

\paragraph*{}
  The $\gamma$-ray singles and  $\gamma - \gamma$ coincidence measurements were carried out at Lawrence Berkeley National Laboratory using the $8\pi$ spectrometer \cite{Martin1987b}, an array of 20 Compton-suppressed Ge detectors.
  Details of the experimental set-up can be found in \cite{Kulp2004a}.
  Gamma-ray singles and $\gamma - \gamma$ coincidence events were recorded, concurrently, event-by-event on magnetic tape in runs lasting 252 h for $\eug$ and 85 h for $\eum$.  
  Single-detector events were scaled down by accepting only one event out of every 128 (24) of these events in the trigger logic for the $\eug$ ($\eum$) experiments.
  This was done to reduce dead time in the data acquisition system so that coincidence information was maximized.

\section{\label{analysis}DATA ANALYSIS}

\paragraph*{}
  Data tapes were scanned (as described in \cite{Kulp2004a}) to provide $\gamma$-ray singles and $\gamma - \gamma$ coincidence spectra.
  The data obtained contained  $6\times 10^8$ ($2\times 10^7$) $\gamma-\gamma$ coincidence events and $2\times 10^9$ ($2\times 10^8$) singles events for the $\eug$ ($\eum$) experiments.
  The $\eug$ source contained 0.8$\%$ $\euz$, and the $\eum$ sources contained
1.4$\%$ $\eug$ and 0.01$\%$ $\euz$, determined as decay rates in this study.

\paragraph*{}
  Calibration for energies and intensities of lines in the $\eu$ decay was achieved internally, i.e., use was made of the fact that the strong lines in the $\eug$ decay serve \cite{Artna-Cohen1996} as a 
secondary $\gamma$-ray energy and intensity calibration source.  
  Peak areas were fitted in all spectra using the program {\sc gf3} \cite{Radware}.

\paragraph*{}
 Summing corrections to peak areas were considered and found to be smaller than the deduced uncertainties in measured $\gamma$-ray intensities.
  Losses from summing were typically on the order of $0.50\%$.
  The largest summing gain identified was for the weak ($I_\gamma = 0.0010\,(4)$) $1757 \rightarrow 122$, 1635~keV transition in the decay of $\eug$;  this was a $10\%$ effect resulting from coincidence summing in the $\sum 671+964$, $\sum 523+1112$, and $\sum 1390+245$ cascades.
  Other summing gains were similarly smaller than uncertainties reported for the measured $\gamma$-ray intensities in this work.

\paragraph*{}
  The long-term energy resolution of singles spectra, summed over all 20 detectors, was 1.8 keV (1.7 keV) at 122 keV and 3.3 keV (3.1 keV) at 1112 keV for the $\eug$ ($\eum$) experiments.
  Energy calibration for both decay studies was made with a linear fit corrected with a cubic nonlinearity function describing keV/channel using adopted \cite{Artna-Cohen1996} values for input.
  Forty-one lines were used in the energy calibration of the $\eug$ decay spectra, while 47 lines were used in the $\eum$ study.
  In both decays, the systematic uncertainty in the energy calibration is deduced to be $\pm 0.14$~keV.

\paragraph*{}
  The $\eug$ $\gamma$-ray singles efficiency calibration was made with a polynomial (containing terms up to quartic) describing log(efficiency) versus log(energy), fitted to 10 
strong lines in the $\eug$ decay (122, 245, 344, 411, 779, 867, 964, 1112, 1213, and 1408~keV). 
  A systematic uncertainty of $2.0\%$ in this 
efficiency curve is deduced by comparing the calculated intensities of 
28 precisely-known (uncertainty $< 2\%$) lines in the $\eug$ decay with the 
adopted \cite{Artna-Cohen1996} intensity values.
  Thirteen $\gamma$ rays from $\eug$ decay (296, 368, 411, 779, 867, 1086, 1090, 1112, 1213, 1299, 1408, 1458, 1528) were used to make a linear fit describing log(efficiency) versus log(energy) for the $\eum$ $\gamma$-ray singles spectrum.   
  A deduced systematic uncertainty for the $\eum$ singles efficiency fit is $\pm 5\%$.

\paragraph*{}
  A 30 ns coincidence time gate was used to construct a prompt $\gamma-\gamma$ coincidence matrix for the $\eug$ decay study, while a 25 ns gate was used in the $\eum$ study.
  Random coincidences were removed by subtracting a background matrix constructed using a delayed-time gate of the same coincidence time width as the prompt matrix and scaled down to remove known accidental coincidences in the final matrix.
  
\paragraph*{}
  The short coincidence times used to construct the $\gamma-\gamma$ coincidence matrices, while effectively removing random coincidences, potentially invalidates the efficiency calibrations made with the singles spectra when used for the coincidence data.
  A polynomial coincidence efficiency curve (containing terms up to quartic) describing log(efficiency) versus log(energy) was constructed for both matrices using an iterative process described as follows.
  
\paragraph*{}
  The number of coincidence counts, $N_{12}$, observed between a pair of $\gamma$ rays in direct cascade can be expressed (cf. Eqs. 4 and 7 in \cite{Wapstra1965}) as
\begin{equation}
\label{Ndef}
  N_{12} = N I_{\gamma_1} \varepsilon_{\gamma_1} B_{\gamma_2} \varepsilon_{\gamma_2} \varepsilon_{12} \eta(\theta_{12});
\end{equation}
where $N$ is a number characterizing the coincidence data for a given decaying isotope, $I_{\gamma_1}$ is the intensity of the ``feeding'' $\gamma$ ray of the pair, $B_{\gamma_2}$ is the branching fraction of the ``draining'' $\gamma$ ray of the pair, $\varepsilon_{\gamma_1}$ and $\varepsilon_{\gamma_2}$ are singles photopeak efficiencies, $\varepsilon_{12}$ is the coincidence efficiency (the factor that distorts the singles efficiency), and $\eta(\theta_{12})$ is the angular correlation attenuation factor.
  A selection of coincident pairs, listed in Table \ref{coincidence calibration}, was used to determine the coincidence efficiency curves.
  First, the normalization constant $N$ was computed using fitted areas in coincidence spectra and the photopeak efficiency calculated from the appropriate singles efficiency curve (assuming negligible angular correlations, discussed below).
  Next, values of $\varepsilon_{\gamma_1}$ and $\varepsilon_{\gamma_2}$ were varied to minimize the spread in $N$.
  The adjusted efficiency values, essentially products of $\varepsilon_{12} $ and $\varepsilon_{\gamma_1}$ or $\varepsilon_{\gamma_2}$, respectively, were then plotted and fitted with a polynomial to produce a new iteration of the coincidence efficiency curve.
  This process was continued until it  converged on a self-consistent average value of $N$.
  In the $\eug$ decay study, for example, the standard deviation of calculated $N$ values for the  coincidence pairs in Table \ref{coincidence calibration} is $1.68\%$ of the mean.
  
\begin{table}[htb]
\caption{\label{coincidence calibration} Coincident pairs of $\gamma$-ray transitions in the decay of $\eug$ used for coincidence efficiency calibration.  Pairs marked with a star were not used in the $\eum$ decay calibration because the parent levels are populated in both $\eug$ and $\eum$ decays.
Data are taken from \cite{Artna-Cohen1996}.}
\begin{ruledtabular}
\begin{tabular}{D{-}{-}{4}D{.}{.}{3}D{-}{-}{4}D{.}{.}{3}D{-}{-}{4}D{.}{.}{3}}
 \multicolumn{1}{c}{$\gamma_1-\gamma_2$} &  \multicolumn{1}{c}{$I_{\gamma_1}B_{\gamma_2}$} &  
 \multicolumn{1}{c}{$\gamma_1-\gamma_2$} &  \multicolumn{1}{c}{$I_{\gamma_1}B_{\gamma_2}$} &  
 \multicolumn{1}{c}{$\gamma_1-\gamma_2$} &  \multicolumn{1}{c}{$I_{\gamma_1}B_{\gamma_2}$}\\   
\hline
 \multicolumn{6}{c}{Data used for both $\eug$ and $\eum$ decays} \\
245-122 &  13.06 & 444-245^\star & 1.10  & 411-344 & 8.10   \\
689-122^\star &  1.47   & 867-245 & 14.31  & 586-344^\star & 1.67   \\
919-122 &  0.73   & 1005-245 & 2.18  & 779-344 & 46.93   \\
964-122 &  25.14   & 1213-245 & 4.79  & 1090-344 & 6.26   \\
1112-122 &  23.49   & & & 1299-344 & 5.89   \\
1408-122 &  36.16   & & & &   \\
 \multicolumn{6}{c}{Additional data for $\eum$ decay} \\
563-122 & 0.71 & & & 703-344 & 0.45 \\
841-122 & 46.08 & & & 970-344 & 3.97 \\
1389-122 & 2.43 & & & 1412-344 & 0.30 \\
1559-122 & 0.025 & & & &
\end{tabular}
\end{ruledtabular}
\end{table}  

\paragraph*{}
  Negligible angular correlation attenuation factors can be assumed for the efficiency curve calculations because the icosahedral symmetry of the $8\pi$ detector array results in minimal angular correlation attenuation factors.
  Due to its symmetry, the array possesses a unique property, namely, the sum
\begin{equation}
  \frac{\sum_\theta n(\theta) P_\lambda(\cos{\theta})}{\sum_\theta n(\theta)}  \approx 0, \; \lambda = 2,4,
\end{equation}
where $\theta =   41.8^\circ$, $70.5^\circ$, $109.5^\circ$, $138.2^\circ$, $180.0^\circ$ is the angle between a detector pair and $n(41.8^\circ) = n(138.2^\circ) = 60$, $n(70.5^\circ) = n(109.5^\circ) = 120$, and $n(180.0^\circ) = 20$ are the combinations of detector pairs with an angle $\theta$ between the detectors.
  The largest angular correlation attenuation factor occurs for a spin $0 \rightarrow 2 \rightarrow 0$ cascade where $\eta(\theta_{12}) = 0.92430$, i.e., a $7.57\%$ effect.
  Other coincidence cascades result in angular correlation distortions $\le 1.86\%$.
  No $0 \rightarrow 2 \rightarrow 0$ cascades were used in the coincidence efficiency calibration.
 
\paragraph*{}
  The intensities of 38 well-known $\gamma$ rays in the decay of $\eug$ were calculated using Eq. \ref{Ndef} and the fitted areas in coincidence gates; these values were then compared with adopted values \cite{Artna-Cohen1996} to determine a systematic uncertainty for the efficiency curve.  
  We deduce an uncertainty of $\pm 3.3\%$ in the coincidence efficiency curve for the $\eug$ decay study based upon this comparison.
  A $\pm 5.5\%$ uncertainty in the efficiency curve is similarly determined for the $\eum$ decay study.

\paragraph*{}
  Coincidence measurements ``from below'' are favored due to the product of $I_{\gamma_1}  B_{\gamma_2}$ in Eq. \ref{Ndef}.
  Remarkably weak $\gamma$ rays can be observed in this way if most of the decay out of a level goes through one draining transition (e.g., $B_{\gamma} \sim$1 for the $0^+_2 \rightarrow 2^+_1$ 562.98 keV transition discussed in Section \ref{ground state decay}). 
    While the branching fractions, $B_{\gamma_2}$, require detailed knowledge of the decay transitions from the levels for which draining intensities are selected, these values are well known in $\sm$ from previous studies.
  As can be seen in Table \ref{coincidence gates}, our measurements typically agree with $B_\gamma$ values deduced from adopted \cite{Artna-Cohen1996} $I_\gamma$ values within the $3.3\%$ uncertainty in the coincidence efficiency curve.  
  Further, in cases where a single $\gamma$ ray is the only decay out of a level, the accuracy of $B_{\gamma}$ is limited only by the precision in the calculated (or measured) internal conversion coefficient.
  In the case of the three gating transitions used for calibration (cf. Table \ref{coincidence calibration}), the $B_{\gamma}$ values for the 122, 245, and 344 keV $\gamma$ rays are known to $0.8\%$, $0.1\%$, and $0.1\%$ precision, respectively, based upon the precision in the theoretical internal conversion coefficient calculations \cite{Kibedi2005}.
  
\begin{table}[htb]
\caption{\label{coincidence gates} Branching fractions for $\gamma$-ray transitions used to measure coincidence intensities in the decay of $\eug$.  The $B_{\gamma}$ (NDS) values are calculated based upon the evaluated $I_\gamma$ values in the  ``Adopted Levels, Gammas'' for $\sm$ in \cite{Artna-Cohen1996}.  Comments indicate:  (1) $B_\gamma$ value is a function only of $\alpha$ for the $\gamma$-ray transition, (2) the $B_{\gamma}$ value determined in this study, rather than the $B_{\gamma}$ (NDS) value, was used to calculate $I_\gamma$ for coincident $\gamma$ rays.}
\begin{ruledtabular}
\begin{tabular}{D{.}{.}{2}D{.}{.}{3}D{.}{.}{4}D{.}{.}{4}D{.}{.}{3}c}
 \multicolumn{1}{c}{$E_i$} &  \multicolumn{1}{c}{$E_{\gamma}$} &  \multicolumn{1}{c}{$B_{\gamma}$} &  \multicolumn{1}{c}{$B_{\gamma}$ (NDS)} & \multicolumn{1}{c}{$\%$ change}   & Comments \\
\hline
121.8   &  121.71 & 0.461 & 0.461 & & 1  \\
366.5   &  244.72 & 0.903 & 0.903 & & 1 \\
684.7   &  562.98 & 0.974 & 0.977 & -0.23\% &  \\
706.9   &  340.37 & 0.963 & 0.963 & & 1 \\
810.5   &  688.67 & 0.562 & 0.542 &  3.53\%  & 2 \\
             &  810.48 & 0.212 & 0.202 &  4.93\%  & 2 \\
963.4   &  841.61 & 0.557 & 0.548 &  1.60\%  & \\
             &  963.41 & 0.439 & 0.451 & -2.55\%  & \\
1023.0 &  656.51 & 0.528 & 0.564 & -6.33\%  & 2 \\
              &  901.34 & 0.315 & 0.309 &  1.94\%  & 2 \\
1041.1 &  674.64 & 0.292 & 0.287 &  1.48\%  & \\
              &  919.41 & 0.707 & 0.711 & -0.63\%  & \\
1082.9 &  961.30 & 0.841 & 0.827 &  1.71\%  &  \\
1085.9 &  964.14 & 0.574 & 0.580 & -0.96\%  & \\
              & 1085.87 & 0.415 & 0.405 &  2.43\%  & \\
1221.5 &  514.89 & 0.209 & 0.193 &  8.22\%  & 2 \\
              &  855.36 & 0.786 & 0.805 & -2.45\%  & 2 \\
1233.9 &  867.41 & 0.232 & 0.236 & -1.77\%  & \\
              & 1112.08 & 0.764 & 0.758 &  0.77\%  & \\
1292.8 &  329.43 & 0.189 & 0.189 &  0.00\%  & \\
              &  926.32 & 0.419 & 0.411 &  1.99\%  & \\
              & 1292.76 & 0.155 & 0.156 & -0.52\%  & \\
1371.7 & 1005.24 & 0.735 & 0.742 & -1.03\%  & \\
              & 1249.95 & 0.220 & 0.216 &  1.89\%
\end{tabular}
\end{ruledtabular}
\end{table}

\paragraph*{}
  Coincidence intensities were measured for all observed $\gamma$ rays in the decay of $\eug$ and $\eum$, except where no coincidence measurement was possible (e.g., the 1769.09 keV $\gamma$ ray to the ground state from the $2^+$ level at 1769.1 has no $\gamma$ ray in coincidence).
  In the case of a $\gamma$ ray which depopulates a level directly to the ground state, a relative intensity was determined through a coincidence gate on a $\gamma$ ray feeding the level, then normalized to another $\gamma$ ray out of the level.
  Where possible, coincidence gates on at least two transitions fed by a given $\gamma$ ray have been used to determine the $\gamma$-ray intensity.
  Table \ref{coincidence gates} lists the $B_\gamma$ values for transitions used to measure $\gamma$-ray intensities in both the $\eug$ and $\eum$ decays.
  Values of $B_\gamma$ deduced from adopted \cite{Artna-Cohen1996} $I_\gamma$ values were used, except for the three levels (i.e., the 810.5, 1023.0, 1221.5~keV levels) where our measured $B_\gamma$ disagreed with these values by more than our deduced uncertainty.
 
\section{\label{results}EXPERIMENTAL RESULTS} 

\paragraph*{}
  Both the decay of $\eug$ and $\eum$ to $\sm$ were studied in detail to examine the low-energy low-spin structure of $\sm$.
  Results from the $\eug$ decay study are presented in Section \ref{ground state decay} and results from the $\eum$ study are presented in Section \ref{isomer results}.
  From a comparison of these experimental results with complementary $(n,n'\gamma)$ and $(\alpha,2n\gamma)$ studies \cite{Garrett2005}, we deduce that all possible $E2$ transitions between states below 1550 keV with transition energies $> 130$ keV are observed.
  
\subsection{Transitions assigned to the decay of $\eug$ and coincidence intensities}
\label{ground state decay}

\paragraph*{}
  Measured energies and intensities for $\gamma$ rays assigned to the decay of $\eug$ are listed in Table \ref{longlived}.
  Reported $\gamma$-ray intensities, $I_\gamma$, are normalized relative to the 344~keV line in the decay of $\eug \rightarrow \gd$, where $I_\gamma(344) \equiv 100$, as per standard practice \cite{Artna-Cohen1996} for this decay.
  New assignments have been made on the basis of coincidence spectroscopy, and all transitions have been measured in coincidence where practicable.

\paragraph*{}
  A total of 121 $\gamma$-ray transitions are assigned to the $\eug \rightarrow \sm$ decay scheme.
   Of these transitions, 42 $\gamma$ rays were previously unassigned or unobserved in the evaluated decay scheme.
  Four levels in $\sm$ (at 1082.9, 1221.5, 1776.2, and 1821.2~keV), known from other spectroscopic probes, are established for the first time as populated in the decay of $\eug$.
  We confirm Barrette et al.'s report \cite{Barrette1971} of population of the $4^-$ level at 1682.1~keV which, while known from other spectroscopic probes, was not indicated in {\em Nuclear Data Sheets} \cite{Artna-Cohen1996} as populated in the decay of $\eug$.
  A level in $\sm$ at 1779~keV is reported for the first time in the present study (we have confirmed this level in an $(n,n'\gamma)$ study).

\begin{ruledtabular}
\begin{longtable}{rD{.}{.}{6}D{.}{.}{8}c}
\caption{\label{longlived} Transitions in $\sm$ observed in the decay of $\eug$.  The intensity of the 344 keV line in $\gd$ ($I_\gamma \equiv 100$) is used for normalization.  Notes indicate: (1) $I_\gamma$ determined from the singles spectrum, (2) $I_\gamma$ determined from a coincidence gate above the transition and normalized to the strongest $\gamma$ ray out of the level, (3) a level newly assigned to the decay scheme as a result of this work, (4) $I_\gamma$ determined from a coincidence gate above the transition and normalized to $I_\gamma$(272), (5) a $\gamma$ ray newly assigned as a result of this work, and (6) reported $I_\gamma$ is an upper limit calculated for an unobserved $\gamma$ ray.}\\
 \multicolumn{1}{l}{$E_i$ (keV) \qquad $J^{\pi}$ \hfill}  &  \multicolumn{2}{c}{$I_{tot}^{in}$ / $I_{tot}^{out}$} & \\
 \multicolumn{1}{r}{$E_f$ (keV)} & \multicolumn{1}{c}{$E_{\gamma}$ (keV)} & \multicolumn{1}{c}{$I_{\gamma}$} & notes  \\
 \hline \\
\endfirsthead
 \multicolumn{1}{l}{$E_i$ (keV) \qquad $J^{\pi}$ \hfill} &  \multicolumn{2}{c}{$I_{tot}^{in}$ / $I_{tot}^{out}$}  & \\
 \multicolumn{1}{r}{$E_f$ (keV)} & \multicolumn{1}{c}{$E_{\gamma}$ (keV)} & \multicolumn{1}{c}{$I_{\gamma}$} & notes  \\
 \hline \\
\endhead
  \multicolumn{4}{c}{}\\ 
 \multicolumn{1}{l}{121.71\,(14) \quad $2^{+}$ \hfill}  & \multicolumn{2}{c}{226 (8) / 232 (5)} & \\ 
   0.0 &  121.71\,(14) & 106.8\,(21) & 1\\  
  \multicolumn{4}{c}{}\\ 
 \multicolumn{1}{l}{366.43\,(14) \quad $4^{+}$ \hfill} & \multicolumn{2}{c}{28.0 (9) / 31.8 (10)} & \\ 
 121.8 &  244.72\,(14) & 28.7\,(9) & \\  
  \multicolumn{4}{c}{}\\ 
 \multicolumn{1}{l}{684.69\,(14) \quad $0^{+}$ \hfill} & \multicolumn{2}{c}{0.091 (15) / 0.078 (7)} & \\ 
 121.8 &  562.98\,(14) & 0.076\,(7) & \\  
  \multicolumn{4}{c}{}\\ 
 \multicolumn{1}{l}{706.94\,(14) \quad $6^{+}$ \hfill} & \multicolumn{2}{c}{0.115 (14) / 0.105 (3)} & \\ 
 366.5 &  340.51\,(14) & 0.101\,(3) & \\  
  \multicolumn{4}{c}{}\\ 
 \multicolumn{1}{l}{810.41\,(14) \quad $2^{+}$ \hfill} & \multicolumn{2}{c}{1.35 (5) / 5.84 (19)} & \\ 
 684.7 &  125.66\,(14) & 0.0193\,(7) & \\  
 366.5 &  444.01\,(17) & 1.12\,(4) & \\  
 121.8 &  688.67\,(14) & 3.28\,(11) & \\  
   0.0 &  810.48\,(14) & 1.24\,(4) & 2 \\  
  \multicolumn{4}{c}{}\\ 
 \multicolumn{1}{l}{963.33\,(14) \quad $1^{-}$ \hfill} & \multicolumn{2}{c}{1.16 (4) / 1.22 (4)} & \\ 
 121.8 &  841.61\,(14) & 0.679\,(23)  & \\  
   0.0 &  963.41\,(15) & 0.536\,(17)  & 2 \\  
  \multicolumn{4}{c}{}\\ 
 \multicolumn{1}{l}{1022.93\,(14) \quad $4^{+}$ \hfill} & \multicolumn{2}{c}{0.149 (9) / 1.04 (4)} & \\ 
 810.5 &  212.39\,(14) & 0.0814\,(27)  & \\  
 706.9 &  316.08\,(17) & 0.038\,(9)  & \\  
 366.5 &  656.51\,(14) & 0.549\,(18)  & \\  
 121.8 &  901.34\,(15) & 0.328\,(13)  & \\  
  \multicolumn{4}{c}{}\\ 
 \multicolumn{1}{l}{1041.11\,(14) \quad $3^{-}$ \hfill} & \multicolumn{2}{c}{2.00 (9) / 2.25 (7)} & \\ 
 366.5 &  674.64\,(14) & 0.656\,(21)  & \\  
 121.8 &  919.41\,(14) & 1.59\,(5)  & \\  
  \multicolumn{4}{c}{}\\ 
 \multicolumn{1}{l}{1082.43\,(15) \quad $0^{+}$ \hfill} & \multicolumn{2}{c}{0.0284 (12) / 0.029 (4)} & 3 \\ 
 963.4 &  118.97\,(15) & 0.0020\,(4)  & 4, 5 \\  
 810.5 &  272.47\,(18) & 0.00205\,(25)  & \\  
 121.8 &  961.3\,(3) & 0.024\,(3)  & 4, 5 \\  
  \multicolumn{4}{c}{}\\ 
 \multicolumn{1}{l}{1085.87\,(14) \quad $2^{+}$ \hfill} & \multicolumn{2}{c}{13.4 (5) / 93 (3)} & \\ 
 810.5 &  275.37\,(14) & 0.137\,(5)  & \\  
 684.7 &  401.38\,(16) & 0.00255\,(21)  & 5 \\  
 366.5 &  719.39\,(30) & 0.94\,(3)  & \\  
 121.8 &  964.14\,(14) & 53.4\,(17)  & \\  
   0.0 & 1085.87\,(14) & 38.6\,(13)  & 2 \\  
  \multicolumn{4}{c}{}\\ 
 \multicolumn{1}{l}{1221.82\,(16) \quad $5^{-}$ \hfill} & \multicolumn{2}{c}{0.0089 (4) / 0.0098 (13)} & 3 \\ 
 706.9 &  514.89\,(17) & 0.00205\,(19)  & 5 \\  
 366.5 &  855.36\,(19) & 0.0077\,(11)  & 5 \\  
  \multicolumn{4}{c}{}\\ 
 \multicolumn{1}{l}{1233.80\,(14)\quad $3^{+}$ \hfill} & \multicolumn{2}{c}{2.18 (10) / 67.3 (22)} & \\ 
1085.9 &  148.00\,(15) & 0.077\,(4)  & \\  
1023.0 &  210.95\,(14) & 0.0143\,(6)  & 5 \\  
 810.5 &  423.40\,(14) & 0.0112\,(8)  & \\  
 366.5 &  867.41\,(14) & 15.7\,(5)  & \\  
 121.8 & 1112.08\,(14) & 51.5\,(17)  & \\  
  \multicolumn{4}{c}{}\\ 
 \multicolumn{1}{l}{1292.75\,(14) \quad $2^{+}$ \hfill} & \multicolumn{2}{c}{0.074 (5) / 2.36 (8)} & \\ 
1085.9 &  207.03\,(23) & 0.0043\,(9)  & 5 \\  
1082.9 &  209.84\,(14) & 0.0241\,(10)  & 5 \\  
1041.1 &  251.65\,(14) & 0.247\,(9)  & \\  
1023.0 &  269.75\,(14) & 0.0285\,(25)  & \\  
 963.4 &  329.43\,(14) & 0.446\,(15)  & \\  
 810.5 &  482.35\,(14) & 0.093\,(3)  & \\  
 684.7 &  608.06\,(15) & 0.0010\,(3)  & 5 \\  
 366.5 &  926.32\,(14) & 0.99\,(3)  & \\  
 121.8 & 1171.20\,(20) & 0.141\,(7)  & \\  
   0.0 & 1292.76\,(14) & 0.366\,(12)  & 2 \\  
  \multicolumn{4}{c}{}\\ 
 \multicolumn{1}{l}{1371.69\,(14) \quad $4^{+}$ \hfill}  & \multicolumn{2}{c}{0.0572 (25) / 3.24 (12)} & \\ 
1233.9 &  137.56\,(22) & 0.0028\,(5)  & 5 \\  
1221.5 &  150.20\,(14) & 0.00370\,(24)  & 5 \\  
1085.9 &  285.98\,(18) & 0.037\,(6)  & \\  
1041.1 &  330.67\,(14) & 0.0377\,(19)  & \\  
1023.0 &  348.86\,(16) & 0.0052\,(8)  & 5 \\  
 810.5 &  561.26\,(17) & 0.0058\,(5)  & \\  
 706.9 &  664.57\,(20) & 0.0373\,(19)  & \\  
 366.5 & 1005.24\,(15) & 2.38\,(8)  & \\  
 121.8 & 1249.95\,(14) & 0.713\,(24)  & \\  
  \multicolumn{4}{c}{}\\ 
 \multicolumn{1}{l}{1529.77\,(14) \quad $2^{-}$ \hfill} & \multicolumn{2}{c}{no $\gamma$ in / 94 (4)} & \\ 
1292.8 &  237.02\,(14) & 0.0245\,(13)  & 5 \\  
1233.9 &  295.99\,(14) & 1.64\,(7)  & \\  
1085.9 &  444.02\,(14) & 11.0\,(4)  & \\  
1041.1 &  488.74\,(14) & 1.59\,(7)  & \\  
 963.4 &  566.41\,(14) & 0.514\,(17)  & \\  
 810.5 &  719.36\,(14) & 0.357\,(12)  & \\  
 121.8 & 1408.04\,(14) & 78.5\,(26)  & \\  
  \multicolumn{4}{c}{}\\ 
 \multicolumn{1}{l}{1579.39\,(14)\quad $3^{-}$ \hfill} & \multicolumn{2}{c}{no $\gamma$ in / 7.93 (26)} & \\ 
1371.7 &  207.55\,(17) & 0.0280\,(11)  & \\  
1292.8 &  286.3\,(3) & 0.0084\,(15)  & 5 \\  
1233.9 &  345.91\,(15) & 0.035\,(3)  & 5 \\  
1085.9 &  493.63\,(14) & 0.114\,(5)  &  \\  
1041.1 &  538.27\,(16) & 0.0164\,(12)  & \\  
1023.0 &  556.48\,(15) & 0.0647\,(29)  & \\  
 963.4 &  616.25\,(24) & 0.034\,(3)  & \\  
 810.5 &  769.00\,(14) & 0.433\,(14)  & \\  
 366.5 & 1212.94\,(14) & 5.32\,(17)  & \\  
 121.8 & 1457.68\,(14) & 1.88\,(6)  & \\  
  \multicolumn{4}{c}{}\\ 
 \multicolumn{1}{l}{1613.00\,(14) \quad $4^{+}$ \hfill} & \multicolumn{2}{c}{no $\gamma$ in / 0.077 (5)} & \\ 
1371.7 &  241.0\,           &  <0.0014      & 6 \\
1292.8 &  320.37\,(15) & 0.0074\,(3)  & 5 \\  
1233.9 &  378.15\,(24) & 0.00089\,(21)  & 5 \\  
1221.5 &  391.13\,(14) & 0.0052\,(2)  & 5 \\  
1085.9 &  526.9        & < 0.0007 & 6 \\  
1041.1 &  571.83\,(14) & 0.0167\,(6)  & \\  
1023.0 &  590.13\,(16) & 0.0049\,(3)  & 5 \\  
 810.5 &  802.0\,(5) & 0.00155\,(18)  & 5 \\  
 706.9 &  906.31\,(15) & 0.0345\,(15)  & \\  
 366.5 & 1246.34\,(16) & 0.0035\,(5)  & 5 \\  
 121.8 & 1491.4\,(8) & 0.0022\,(10)  & 5 \\  
  \multicolumn{4}{c}{}\\ 
 \multicolumn{1}{l}{1649.86\,(14) \quad $2^{-}$ \hfill} & \multicolumn{2}{c}{no $\gamma$ in / 3.52 (11)} & \\ 
1292.8 &  357.26\,(15) & 0.0235\,(8)  & \\  
1233.9 &  416.07\,(14) & 0.422\,(15)  & \\  
1085.9 &  563.96\,(14) & 1.91\,(6)  & \\  
1041.1 &  609.23\,(22) & 0.0046\,(6)  & 5 \\  
 963.4 &  686.52\,(15) & 0.0762\,(25)  & \\  
 810.5 &  839.40\,(15) & 0.0685\,(23)  & \\  
 121.8 & 1528.15\,(14) & 1.01\,(3)  & \\  
  \multicolumn{4}{c}{}\\ 
 \multicolumn{1}{l}{1682.10\,(14) \quad $4^{-}$ \hfill} & \multicolumn{2}{c}{no $\gamma$ in / 0.0137 (6)} & 3 \\ 
 366.5 & 1315.67\,(14) & 0.0137\,(6)  & 5 \\  
  \multicolumn{4}{c}{}\\ 
 \multicolumn{1}{l}{1730.22\,(14) \quad $3^{-}$ \hfill} & \multicolumn{2}{c}{no $\gamma$ in / 0.208 (17)} & \\ 
1371.7 &  358.61\,(15) & 0.0066\,(3)  & 5 \\  
1233.9 &  496.56\,(24) & 0.022\,(8)  & \\  
1085.9 &  644.48\,(14) & 0.0279\,(26)  & \\  
1023.0 &  707.20\,(17) & 0.0057\,(3)  & 5 \\  
 963.4 &  766.38\,(18) & 0.00263\,(27)  & 5 \\  
 810.5 &  919.89\,(15) & 0.0258\,(11)  & 5 \\  
 366.5 & 1363.82\,(14) & 0.097\,(3)  & \\  
 121.8 & 1608.29\,(18) & 0.0203\,(11)  & \\  
  \multicolumn{4}{c}{}\\ 
 \multicolumn{1}{l}{1757.10\,(14) \quad $4^{+}$ \hfill} & \multicolumn{2}{c}{no $\gamma$ in / 0.215 (17)} & \\ 
1371.7 &  385.41\,(21) & 0.0212\,(9)  & \\  
1292.8 &  464.28\,(14) & 0.0017\,(7)  & 5 \\  
1233.9 &  523.21\,(15) & 0.0575\,(24)  & \\  
1085.9 &  671.06\,(18) & 0.105\,(4)  & \\  
1023.0 &  734.4\,(5) & 0.0040\,(4)  & 5 \\  
 810.5 &  947.15\,(14) & 0.0041\,(7)  & 5 \\  
 706.9 & 1050.1\,(6) & 0.0027\,(11)  & 5 \\  
 366.5 & 1390.65\,(17) & 0.0159\,(6)  & \\  
 121.8 & 1635.38\,(20) & 0.0010\,(4)  & \\  
  \multicolumn{4}{c}{}\\ 
\multicolumn{1}{l}{1769.12\,(14) \quad $2^{+}$ \hfill} & \multicolumn{2}{c}{no $\gamma$ in / 0.317 (27)} & \\ 
1529.8 &  239.4             & < 0.09          & 6 \\
1371.7 &  397.75\,(26) & 0.00140\,(22)  & 5 \\  
1292.8 &  476.40\,(16) & 0.0087\,(8)  & 5 \\  
1233.9 &  535.57\,(15) & 0.0063\,(5)  &  \\
1085.9 &  683.11\,(16) & 0.0165\,(18)  & 5 \\  
1041.1 &  728.10\,(14) & 0.0418\,(14)  & \\  
 963.4 &  805.84\,(14) & 0.0571\,(22)  & \\  
 810.5 &  958.69\,(14) & 0.072\,(5)  & \\  
 684.7 & 1083.96\,(19) & 0.049\,(13)  & 5 \\  
 121.8 & 1647.49\,(25) & 0.0294\,(14)  & \\  
   0.0 & 1769.09\,(14) & 0.0352\,(10)  & 1 \\  
  \multicolumn{4}{c}{}\\ 
\multicolumn{1}{l}{1776.60\,(14) \quad $1^{-},2$ \hfill} & \multicolumn{2}{c}{no $\gamma$ in / 0.0366 (25)} & 3 \\ 
1041.1 &  735.49\,(14) & 0.0186\,(16)  & 5 \\  
 963.4 &  813.29\,(15) & 0.0180\,(9)  & 5 \\  
  \multicolumn{4}{c}{}\\ 
\multicolumn{1}{l}{1779.14\,(14) \quad $2,3,4$ \hfill} & \multicolumn{2}{c}{no $\gamma$ in / 0.0497 (23)} & 3 \\ 
1023.0 &  756.10\,(21) & 0.0196\,(13)  & 5 \\  
 810.5 &  968.74\,(14) & 0.0301\,(10)  & 5 \\  
  \multicolumn{4}{c}{}\\ 
\multicolumn{1}{l}{1821.98\,(23) \quad $3^{+},4^{+},5^{+}$ \hfill} & \multicolumn{2}{c}{no $\gamma$ in / 0.018 (6)} & 3 \\ 
1233.9 &  588.59\,(29) & 0.009\,(4)  & 5 \\  
 366.5 & 1455.1\,(3) & 0.0094\,(18) & 5
\end{longtable}
\end{ruledtabular}

\paragraph*{}
  In cases where a $\gamma$ ray is not (or cannot be) observed in a coincidence gate set on a transition lower in the decay scheme, a coincidence gate on a ``feeding'' transition higher in the decay scheme is used to determine the intensity of the $\gamma$ ray relative to other $\gamma$ rays which depopulate the same level.
  This technique is employed for the intensities of the 810.48, 963.41, 118.97, 961.3, 1085.87, and 1292.76 keV transitions by using, respectively, the measured coincidence intensities of the 688.67, 841.61, 272.47, 272.47, 964.14, and 926.32~keV $\gamma$ rays for normalization.
  Only the intensities of the 121.71 and 1769.09~keV $\gamma$ rays are determined using the singles spectrum. 
  
\begin{figure*}[htb]
\caption{\label{key data} Key data used to assess a recent \cite{Iachello1998, Zamfir1999a, Klug2000, Zamfir2002b} interpretation of $\sm$.
a.  Partial level scheme of $\sm$ indicating the transitions of interest.
b.  The 563 keV $\gamma$-ray-gated $\gamma$-ray coincidence spectrum.
c.  Portions of the $\gamma$-ray spectrum in coincidence with the 689 keV $\gamma$ ray.
d.  Lines observed in coincidence with the 1086 keV $\gamma$ ray.
Coincident $\gamma$ rays are labeled by energy in keV.
A box around the transition energy highlights transitions not placed in the recent evaluations \cite{Artna-Cohen1996, Vanin2004}.
A peak indicated with a filled circle is due to ``spill over'' into the gate from nearby lines.}
\includegraphics[width=17.8 cm]{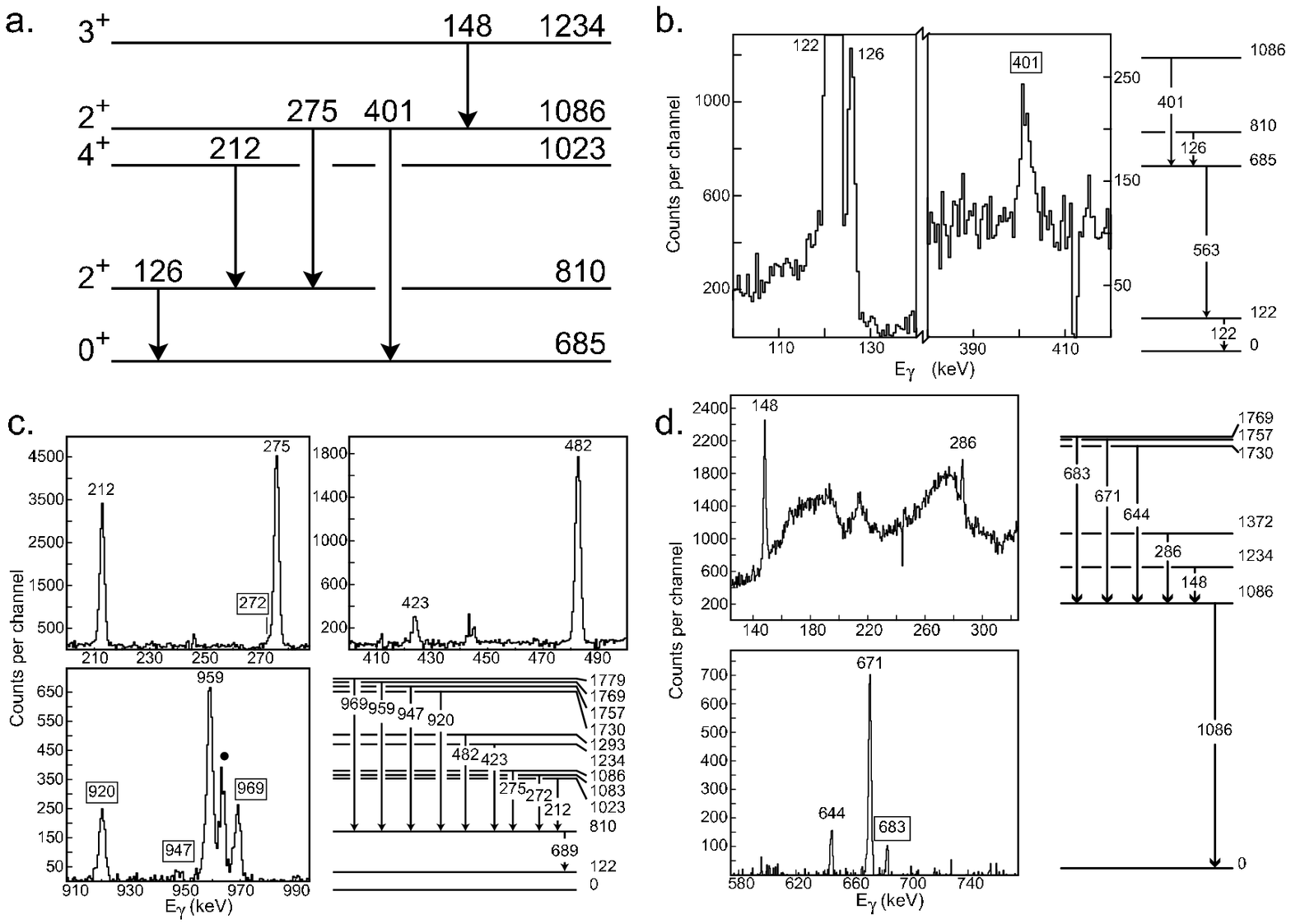}
\end{figure*}

\paragraph*{}
  The partial level scheme given in Fig. \ref{key data}a highlights the principal transitions that stimulated the re-interpretation \cite{Iachello1998, Zamfir1999a} of $\sm$. 
  The recent measurement \cite{Zamfir1999a} of the $2^+_2$ (810) $\rightarrow 0^+_2$ (685) 126 keV transition reported a $B(E2)$ value of 107 (27) W.u., which was interpreted as an indication of a less-collective structure that coexists with the more-collective ground-state band ($B(E2; 2^+_1 \rightarrow 0^+) = 144\,(3)$ \cite{Artna-Cohen1996}).
  In this interpretation \cite{Iachello1998, Casten1998, Zamfir1999a}, the remarkable weakness of the  $2^+_3$ (1086) $\rightarrow 0^+_2$ (685) 401 keV $\gamma$ ray was considered a signature of a forbidden two-phonon transition.
  The  275 keV $\gamma$ ray intensity measured by \cite{Zamfir1999a} indicated that the $2^+_3$ (1086) $\rightarrow 2^+_2$ (810) transition was a collective $E2$, rather than a (non-collective) $M1$ transition as previously reported \cite{Artna-Cohen1996}.
  The $B(E2)$ reported \cite{Klug2000} for the $3^+_1$ (1234) $\rightarrow 2^+_3$ (1086) 148 keV $\gamma$ ray was found to be in accordance with a vibrational prediction, and was argued \cite{Klug2000} to add even further support for the coexistence interpretation.
  A review of our experimental results pertaining to these issues is presented in the following text; while a discussion of the coexistence interpretation, in light of our results, is presented in Section \ref{discussion}.

\paragraph*{}
  The 563 keV $\gamma$-gated $\gamma$-ray spectrum from this study is presented in Fig. \ref{key data}b.
  This gate shows both the $2^+_2$ (810) $\rightarrow 0^+_2$ (685) 126 keV and the $2^+_3$ (1086) $\rightarrow 0^+_2$ (685) 401~keV transitions.
  By measuring these critical transitions through a coincidence gate ``from below'', we have $\sim$7 times higher sensitivity than if we had ``gated from above'' to observe these $\gamma$ rays.
  This sensitivity is crucial for a precise measurement of the 126 keV $\gamma$-ray intensity, which sets the scale for collective models of the structure built on the $0^+_2$ (685) state.
  We determine a relative intensity of $I_\gamma(126) = 0.0193 (7)$, which yields an absolute $B(E2)$ value of 167 (16) W.u. that is slightly stronger than the adopted value \cite{Artna-Cohen1996} of $B(E2) = 144 (3)$ W.u. for the ground-state band $2^+_1$ (122) $\rightarrow 0^+_1$ (0) transition.
  The sensitivity achieved using the 563~keV gate is dramatically illustrated in the observation of the 401~keV $\gamma$ ray, which is remarkable for its extraordinary weakness:  $I_\gamma(401) = 0.00255\,(21)$ (a previous study \cite{Zamfir1999a} could deduce only an intensity upper limit of $<0.008$ for the 401~keV $\gamma$ ray when using a coincidence gate from above).

\paragraph*{}
  The 689 keV $\gamma$-gated $\gamma$-ray spectrum from this study is presented in Fig. \ref{key data}c.
  The $4^+_2$ (1023) $\rightarrow 2^+_2$ (810) 212~keV $\gamma$-ray intensity is a further measure of the collectivity of the structure built on the $0^+_2$ state.
  In addition, an accurate measurement of the $2^+_3$ (1086) $\rightarrow 2^+_2$ (810) 275~keV $\gamma$-ray intensity is critical to determining whether this transition is collective ($E2$) or non-collective ($M1$), as dictated by its previously measured \cite{Goswamy1991} relative internal conversion electron intensity. 

\paragraph*{}
  Four $\gamma$ rays in coincidence with the 689~keV $\gamma$ ray, indicated by boxes around the energy labels on peaks at 272, 920, 947 and 969~keV in Fig. \ref{key data}c, are newly placed in the $\sm$ decay scheme.
  Population of the $0^+_3$ (1083) level and a new level at 1779~keV are inferred, respectively, from the new 272 (and see Fig. \ref{resolved doublets}) and 969~keV $\gamma$ rays observed in coincidence with the 689~keV $\gamma$ ray.

\paragraph*{}
  The 1086 keV $\gamma$-gated $\gamma$-ray spectrum from this study is presented in Fig.~\ref{key data}d.
  The $3^+_1$ (1234) $\rightarrow 2^+_3$ (1086) 148 keV $\gamma$ ray determines the relationship between the $2^+_3$ and $3^+_1$ states.
  Also illustrated in this figure is the new 683 keV $\gamma$ ray depopulating the $2^+$ state at 1769~keV.
      
\begin{figure}[htb]
\caption{\label{unresolved doublets}  A partial level scheme for $\sm$ which shows the $\gamma$-ray and internal conversion intensities required to disentangle the unresolved 444.01-719.36 and 444.02-719.39 doublet pairs.
The total intensities, $I_T$, out of the 810.5 and 1085.9~keV levels are indicated to the right of the respective levels.
The iterative approach adopted to determine the intensities of the 444.01 and 719.39~keV $\gamma$ rays is discussed in the text.
Note that the key 275 keV $\gamma$-ray transition is between the 444.01 and 444.02 $\gamma$ rays; the impact of this is also discussed in Section \ref{eugdecay}.}
\includegraphics[width=8.5cm]{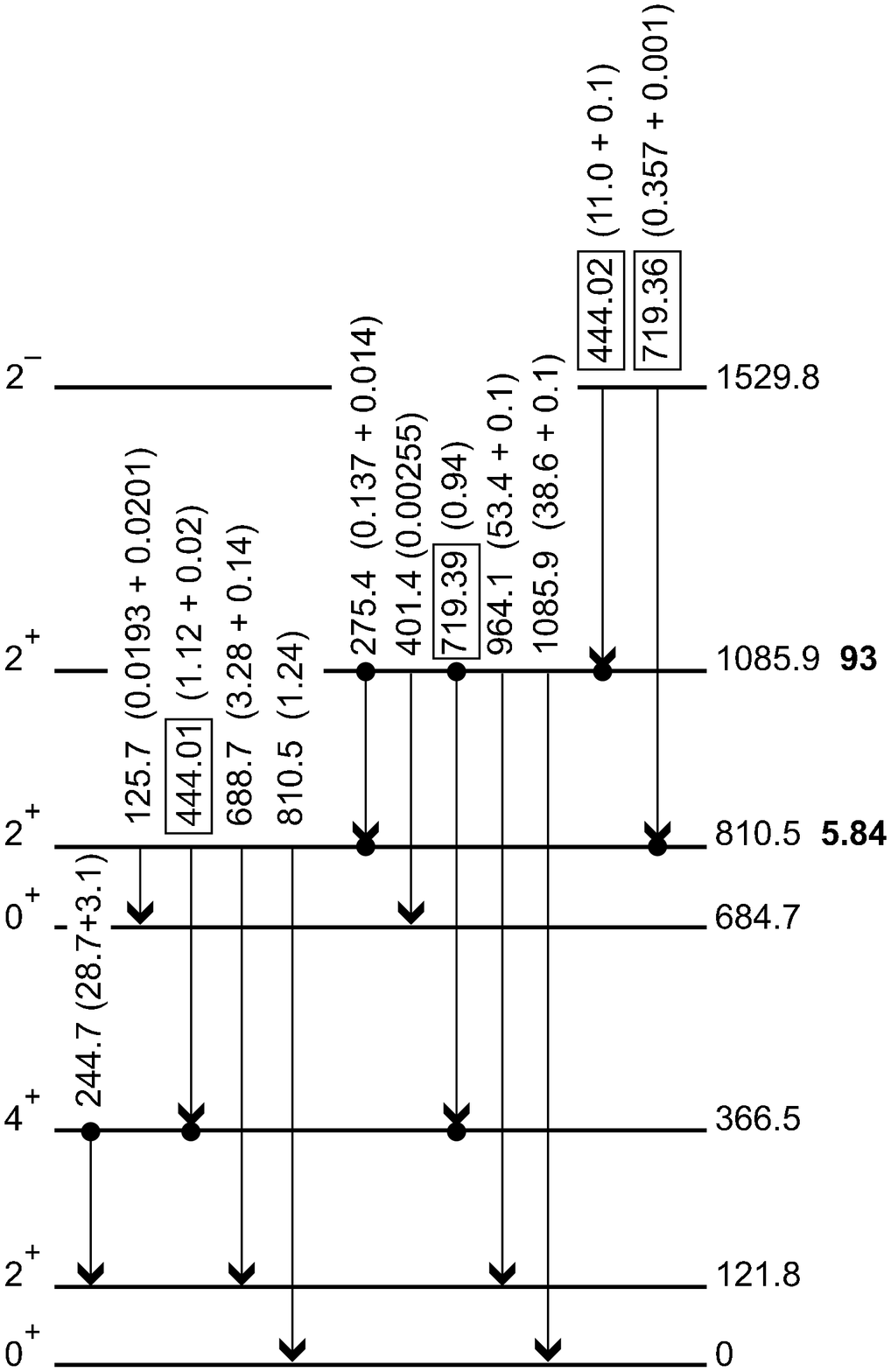}
\end{figure}

\paragraph*{}  
  The decay of $\eug$ has multiple $\gamma$-ray doublets and a number of decay paths which result in pairs of closely-spaced peaks.
  Two of the most troublesome pairs, the 444.01/444.02~keV and 719.36/719.39~keV $\gamma$ rays highlighted in Fig. \ref{unresolved doublets}, are very close in energy and need careful analysis, even when using coincidence intensities.
  The 444.02~keV $\gamma$ ray is observed in the 964.1 and 1085.9~keV $\gamma$-ray gates and the  719.36~keV $\gamma$ ray is observed in the 688.7 and 810.5~keV $\gamma$-ray gates, thus the upper two doublet components may be cleanly measured in coincidence.
  Both of the lower doublet components (the 444.01 and 719.39~keV $\gamma$ rays) feed the level at 366.5~keV and are observed from below only in the 244.7~keV $\gamma$-ray gate, which also will have weak contributions from the upper components (the 444.02 and 719.36~keV $\gamma$ rays). 

\paragraph*{}  
  The number of counts contributing to the 444 and 719~keV peaks in the 245~keV $\gamma$-gated $\gamma$-ray spectrum from the upper doublet components is calculated using a modified version of Equation \ref{Ndef} (cf. Eqs. 4 and 7 in \cite{Wapstra1965})
\begin{equation}
\label{indirect component}
  N_{12}' = N I_{\gamma_1}' \varepsilon_{\gamma_1} B_{\gamma_2} B_{3} \varepsilon_{\gamma_2} \eta(\theta_{1(3)2}),
\end{equation}
where, for the 719~keV doublet, $\gamma_1$ is the upper component (719.36~keV), $\gamma_2$ is the branching fraction of the 245~keV $\gamma$ ray, and $B_{3}$ is the total branching fraction, $B_T$, for the 444.01~keV transition out of the 810.5~keV level.
  For the 444~keV doublet, $\gamma_1$ is the upper 444.02~keV $\gamma$ ray, $B_{\gamma_2}$ is the branching fraction of the 245~keV $\gamma$ ray, and $B_{3}$ is the total branching fraction of the 719.39~keV transition summed with the product $B_T(275.4)\cdot B_T(444.01)$.  
  The angular correlation effects in the indirect cascades are negligible for the summed $8\pi$ array (as discussed in Section \ref{analysis}); and we use $\eta(\theta_{1(3)2}) =1$.
  
\paragraph*{}
  Calculated counts, $N_{12}' $, from the upper doublet components are subtracted from the measured 444 and 719~keV peak areas to determine the coincidence counts, $N_{12}$, for the lower pair of $\gamma$ rays.
  However, because this subtraction alters $B_T(444.01)$ and $B_T(719.39)$, which are used to calculate the $N_{12}' $ values, an iterative approach is required.
  The composite 444~keV peak in the 245~keV $\gamma$-ray-gated spectrum has an intensity of $I_\gamma = 1.24$.
  Using Eq. \ref{indirect component}, a first iteration of the calculation shows that the upper 444.02~keV  contributes $I_\gamma' = 0.12$ to the intensity.
  After subtracting this value and recalculating the $B_T$ of the intervening transitions, the second iteration of the calculation indicates a contribution of $I_\gamma' = 0.11$ from the upper 444.02~keV $\gamma$ ray.
  Both $B_T(444.01)$ and $B_T(719.39)$ converge after two iterations using this method.
    
\begin{figure}[htb]
\caption{\label{resolved doublets} The 210 and 212 keV $\gamma$-ray-gated spectra separate the 270 keV and 272 keV $\gamma$ rays.
Population of the $0^+$ level at 1083 keV, as evidenced by the 272 and 961~keV $\gamma$ rays in the 210~keV gate, was not reported in recent evaluations \cite{Artna-Cohen1996, Vanin2004} of the decay of $\eug$.}
\includegraphics[width=8.5cm]{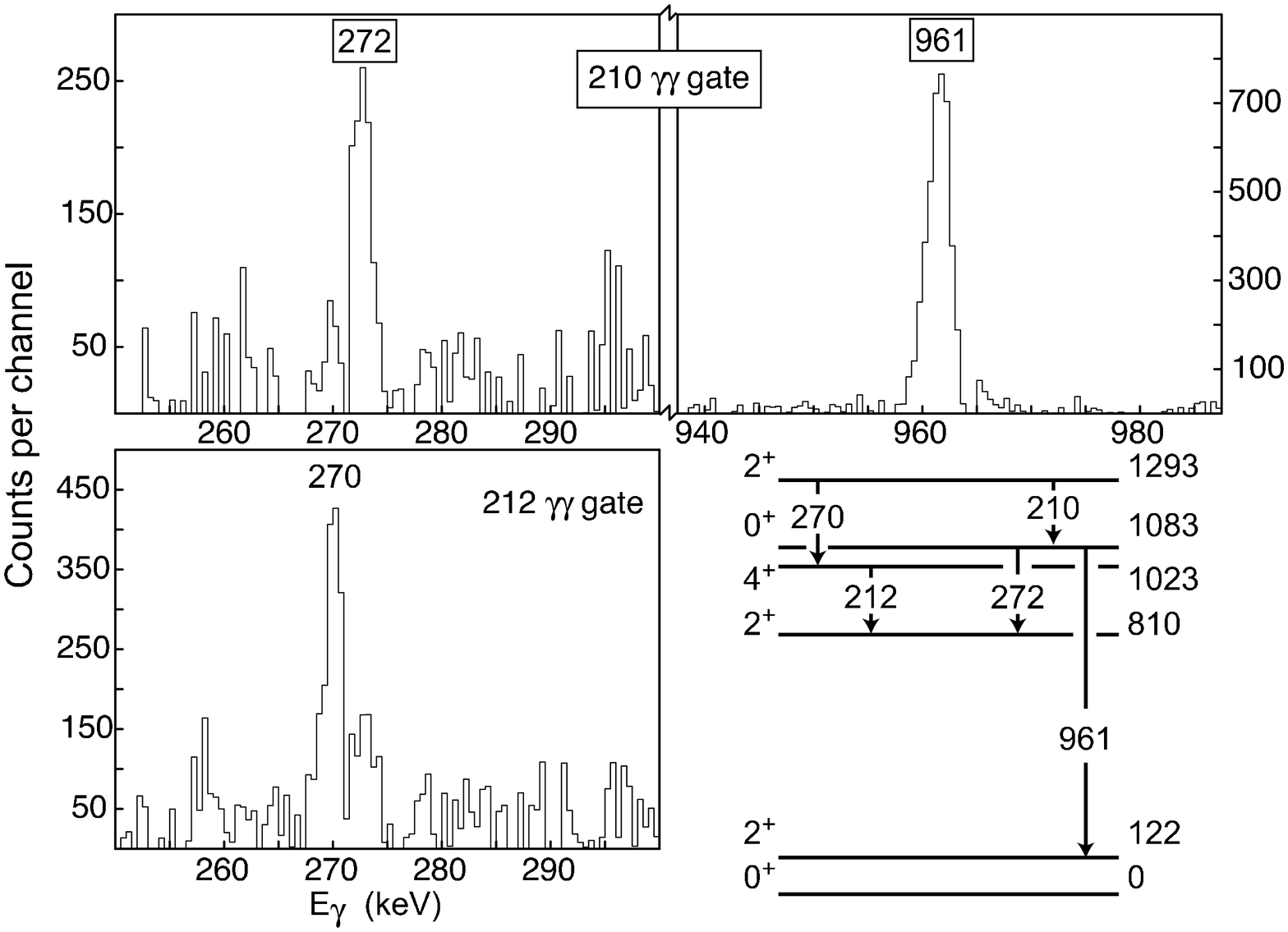}
\end{figure}

\paragraph*{}
  Figure \ref{resolved doublets} shows the coincidences in support of the population of the $0^+_3$ 1083~keV level, which has not been reported previously in $\eug$ (13.6 year) decay studies.
  This figure also illustrates the (counter-intuitive) manifestation of coincidence intensities:  $I_\gamma(272) = 0.00205$, $I_\gamma(270) = 0.0285$; but $B_\gamma (272) I_\gamma (210) = 0.00170$, $B_\gamma (212)I_\gamma (270) = 0.00223$.
  As noted earlier, the 272~keV line intensity is needed to determine the 961~keV line intensity ($I_\gamma(961) = 0.024$), because the latter feeds the $2^+_1$ 122~keV state and is overshadowed by the much stronger $2^+_3$ (1086) $\rightarrow 2^+_1$ (122) 964~keV transition ($I_\gamma (964) = 53.4$).
  
\begin{figure*}[htb]
\caption{\label{new levels} Coincidence spectra illustrating population of levels in addition to those listed in recent evaluations \cite{Artna-Cohen1996, Vanin2004}.
a.  Transitions feeding the $5^-$ 1222 keV level observed in the 855 keV $\gamma$-gated $\gamma$-ray coincidence spectrum.
b.  The 735 keV $\gamma$ ray seen in coincidence with the 919 keV $\gamma$ ray is the strongest transition out of the 1776 keV level.
c.  The 756 keV $\gamma$ ray depopulating a level at 1779 keV is the strongest of several newly-placed  transitions in coincidence with the 657 keV $\gamma$ ray.
d.  The 588 keV $\gamma$ ray in coincidence with the 1112 keV $\gamma$ ray is one of two transitions observed depopulating a level at 1821 keV.
A coincident $\gamma$ ray is labeled by energy in keV, and a line due to spill over from a nearby gate is indicated with a filled circle.
Transitions not placed in the recent evaluations \cite{Artna-Cohen1996, Vanin2004} are highlighted with a box around the transition energy.
A $\Sigma$ label indicates a sum peak.}
\includegraphics[width=17.8 cm]{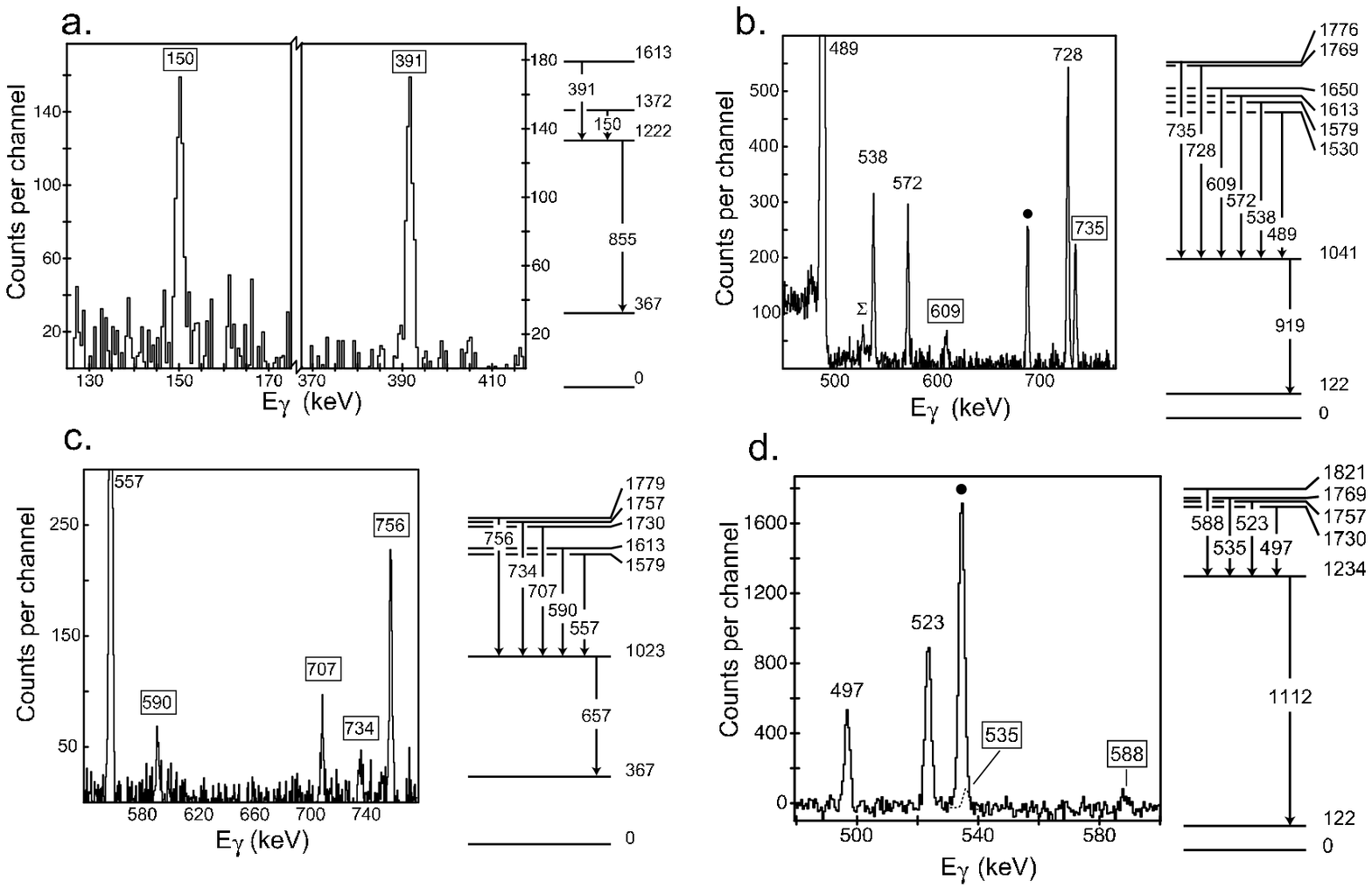}
\end{figure*}
  
\paragraph*{}
  Figure \ref{new levels} shows the coincidences in support of the population of the ($J^\pi_i$ ($E_x$ keV)) levels:  $5^-_1$ (1222); $1^-,2$ (1776); $3^+,4^+,5^-$ (1821).
  These levels in $\sm$ are known \cite{Artna-Cohen1996}, from other spectroscopic probes; but have not been previously observed in the decay of $\eug$.
  The level at 1779~keV, supported by the 756~keV line in Fig. \ref{new levels}c (and the 969~keV line in Fig. \ref{key data}c) is a new level in $\sm$.
  We have confirmed this level and assigned $J^\pi = 3^-$ in an $(n,n'\gamma)$ study \cite{Garrett2005}.
  (We have also assigned a spin parity of $4^-$ to the 1821~keV level, based on $(n,n'\gamma)$ and $(\alpha,2n\gamma)$ studies \cite{Garrett2005}.)

\subsection{\label{isomer results}Transitions assigned to the decay of $\eum$}

\paragraph*{}
  Measured energies and intensities for $\gamma$ rays assigned to the decay of $\eum$ are listed in Table \ref{shortlived}.
    Reported $I_\gamma$ values are normalized relative to the $1^-_1$ (963) $\rightarrow 2^+_1$ (122) 841.58~keV line in the decay of $\eug \rightarrow \sm$, where $I_\gamma(841) \equiv 100$.
  All of these assignments (except the 1510.58 and 1679.96~keV transitions) have been made on the basis of coincidence spectroscopy.
  
\paragraph*{}  
  A total of 32 $\gamma$-ray transitions are assigned to the $\eum \rightarrow \sm$ decay scheme.
  Thirteen transitions are seen for the first time in this decay.
  All of the levels observed to be populated are previously established in the $\sm$ excitation scheme \cite{Artna-Cohen1996}.
  New transitions are observed populating the 1041.1, 1085.9, 1292.8 keV levels.
  Upper limits for unobserved transitions are determined for four transitions.

\begin{ruledtabular}
\begin{longtable}{rD{.}{.}{6}D{.}{.}{8}c}
\caption{\label{shortlived} Transitions in the decay of $\eum$.  The intensity of the 841.58~keV $\gamma$ ray is used for normalization, $I_{\gamma}(841) \equiv 100$.  Comments indicate:  (1) $I_\gamma$ values determined assuming $I_{tot}^{in} = I_{tot}^{out}$ for this level due to $\eug$ contamination, (2)  $I_\gamma$ value corrected for $\eug$ contamination, (3) a newly assigned $\gamma$ ray, (4)  $I_\gamma$ determined through a coincidence gate above the transition and normalized to the strongest $\gamma$ ray out of the level, (5) a level newly assigned to this decay scheme, (6)  $I_\gamma$ calculated using total intensity into level and relative $I_\gamma$ values determined through decay of $\eug$, (7) reported $I_\gamma$ is an upper limit calculated for an unobserved $\gamma$ ray, (8) $I_\gamma$ determined through singles spectrum.}\\
\multicolumn{1}{l}{$E_i$ (keV) \qquad $J^{\pi}$ \hfill}  &  \multicolumn{2}{c}{$I_{tot}^{in}$ / $I_{tot}^{out}$} & \\
\multicolumn{1}{r}{$E_f$ (keV)} & \multicolumn{1}{c}{$E_{\gamma}$ (keV)} & \multicolumn{1}{c}{$I_{\gamma}$} & Comments  \\
\hline \\
\endfirsthead
 \multicolumn{1}{l}{$E_i$ (keV) \qquad $J^{\pi}$ \hfill} &  \multicolumn{2}{c}{$I_{tot}^{in}$ / $I_{tot}^{out}$}  & \\
 \multicolumn{1}{r}{$E_f$ (keV)} & \multicolumn{1}{c}{$E_{\gamma}$ (keV)} & \multicolumn{1}{c}{$I_{\gamma}$} & Comments  \\
 \hline \\
\endhead
   \multicolumn{4}{c}{}\\ 
 \multicolumn{1}{l}{121.84\,(14) \quad $2^{+}$ \hfill}  & \multicolumn{2}{c}{109 (7) /  $\gamma$ ray obsc.} & 1 \\ 
  0.0 &  121.84\,(14) & \multicolumn{1}{c}{$[50]\,(3)$} & 1 \\ 
  \multicolumn{4}{c}{}\\ 
 \multicolumn{1}{l}{366.54\,(14) \quad $4^{+}$ \hfill}  & \multicolumn{2}{c}{0.128 (8) / 0.144 (9)} & \\ 
121.8 &  244.70\,(14) & 0.130\,(8) & 2 \\ 
  \multicolumn{4}{c}{}\\ 
 \multicolumn{1}{l}{684.76\,(14) \quad $0^{+}$ \hfill}  & \multicolumn{2}{c}{0.63 (3) / 1.58 (17)} & \\ 
121.8 &  562.92\,(14) & 1.55\,(17) & 2 \\ 
  \multicolumn{4}{c}{}\\ 
 \multicolumn{1}{l}{810.47\,(14) \quad $2^{+}$ \hfill}  & \multicolumn{2}{c}{0.86 (5) / 0.76 (5)} & \\ 
684.7 &  125.43\,(17) & 0.0025\,(3) & 2, 3\\ 
366.5 &  443.95\,(14) & 0.121\,(8) & 2 \\ 
121.8 &  688.60\,(14) & 0.45\,(3) & 2 \\ 
  0.0 &  810.52\,(14) & 0.162\,(11) & 2, 4 \\ 
  \multicolumn{4}{c}{}\\ 
 \multicolumn{1}{l}{963.42\,(14) \quad $1^{-}$ \hfill}  & \multicolumn{2}{c}{0.161 (13) / 189 (14)} & \\ 
810.5 &  152.77\,(16) & 0.0129\,(11) & \\ 
684.7 &  278.7         & \multicolumn{1}{c}{obscured} & \\ 
121.8 &  841.56\,(14) & \multicolumn{1}{c}{$100\,(6)$} & \\ 
  0.0 &  963.42\,(14) & \multicolumn{1}{c}{$89\,(8)$} & 4 \\ 
  \multicolumn{4}{c}{}\\ 
 \multicolumn{1}{l}{1041.02\,(16) \quad $3^{-}$ \hfill}  & \multicolumn{2}{c}{0.0075 (13) / 0.0071 (9)} & 5 \\ 
366.5 &  674.46\,(16) & 0.00256\,(22) & 2, 3\\ 
121.8 &  919.27\,(24) & 0.0045\,(7) & 2, 3\\ 
  \multicolumn{4}{c}{}\\ 
 \multicolumn{1}{l}{1082.96\,(14) \quad $0^{+}$ \hfill}  & \multicolumn{2}{c}{0.0049 (7) / 1.07 (6)} & \\ 
963.4 &  119.44\,(15) & 0.073\,(6) & 3 \\ 
810.5 &  272.53\,(15) & 0.070\,(4) & \\ 
121.8 &  961.20\,(16) & 0.90\,(5) & \\ 
  \multicolumn{4}{c}{}\\ 
 \multicolumn{1}{l}{[1085.9] \quad $2^{+}$ \hfill} & \multicolumn{2}{c}{0.0036 (12) / $\gamma$ rays obsc.} & 1, 5\\ 
121.8 &     964.1 & [0.002]  & 1, 3, 6 \\ 
     0.0 & 1085.9  & [0.001]  & 1, 3, 6 \\ 
  \multicolumn{4}{c}{}\\ 
 \multicolumn{1}{l}{\em{1289.9} \quad $1,2^{+}$ \hfill}  & \multicolumn{2}{c}{unsupported} & \\ 
684.7 & 605.2 & < 0.0014 & 7 \\ 
121.8 & 1168.2 & < 0.0022 & 7 \\ 
0.0 & 1289.9 & < 0.0011 & 7, 8 \\ 
  \multicolumn{4}{c}{}\\ 
 \multicolumn{1}{l}{[1292.8] \quad $2^{+}$ \hfill} & \multicolumn{2}{c}{0.0051 (8) / $\gamma$ rays obsc.} & 1, 5 \\ 
963.4 & 329.4 & [0.0010] & 1, 3, 6 \\ 
366.5 & 926.3 & [0.0021]  & 1, 3, 6 \\ 
0.0 & 1292.8  & [0.0008]  & 1, 3, 6 \\ 
  \multicolumn{4}{c}{}\\ 
 \multicolumn{1}{l}{1510.80\,(14) \quad $1^{-}$ \hfill}  & \multicolumn{2}{c}{no $\gamma$ in / 6.2 (3) } & \\ 
1292.8 & 218.10\,(15) & 0.0042\,(5) & \\ 
1085.9 & 424.3\,(4) & 0.0012\,(5) & 3 \\ 
1082.9 & 427.9\, & <0.0006 & 7 \\
1041.1 & 469.97\,(20) & 0.0039\,(6) & 3 \\ 
963.4 & 547.36\,(14) & 0.068\,(5) & \\ 
810.5 & 700.28\,(14) & 0.073\,(5) & \\ 
684.7 & 825.99\,(14) & 0.202\,(12) & \\ 
121.8 & 1388.99\,(14) & 5.8\,(3) & \\ 
0.0 & 1510.58\,(14) & 0.0459\,(25) & 8 \\ 
  \multicolumn{4}{c}{}\\ 
 \multicolumn{1}{l}{1680.57\,(14) \quad $1^{-}$ \hfill}  & \multicolumn{2}{c}{no $\gamma$ in / 1.22 (7)} & \\ 
1292.8 & 388.3\,(5) & 0.0008\,(3) & 3 \\ 
1085.9 & 594.7\,(4) & 0.0024\,(7) & 3 \\ 
1082.9 & 597.50\,(14) & 0.0049\,(7) & 3 \\ 
1041.1 & 639.15\,(14) & 0.0036\,(7) & 3 \\ 
  963.4 & 716.84\,(21) & 0.0084\,(13) & 3 \\ 
810.5 & 870.13\,(14) & 0.70\,(4) & \\ 
684.7 & 995.78\,(14) & 0.41\,(2) & \\ 
121.8 & 1558.61\,(15) & 0.056\,(4) & \\ 
0.0 & 1679.96\,(14) & 0.038\,(4) & 8
\end{longtable}
\end{ruledtabular}

\begin{figure*}[htb]
\caption{\label{isomer decay} Modifications to the $\eum$ decay scheme.
a.  An upper limit of $<0.0014$ is calculated for the 605 keV $\gamma$ ray which was previously associated with a 1289 keV level based upon the 563 keV $\gamma$-gated $\gamma$-ray coincidence spectrum.
b.  Gamma rays from the $1^-$ levels at 1511 and 1680 keV are observed in coincidence with the 919 keV $\gamma$ ray that depopulates the $3^-_1$ level at 1041~keV.
The 488 keV $\gamma$ ray is a contaminant from the decay of $\eug$.}
\includegraphics[width=17.8 cm]{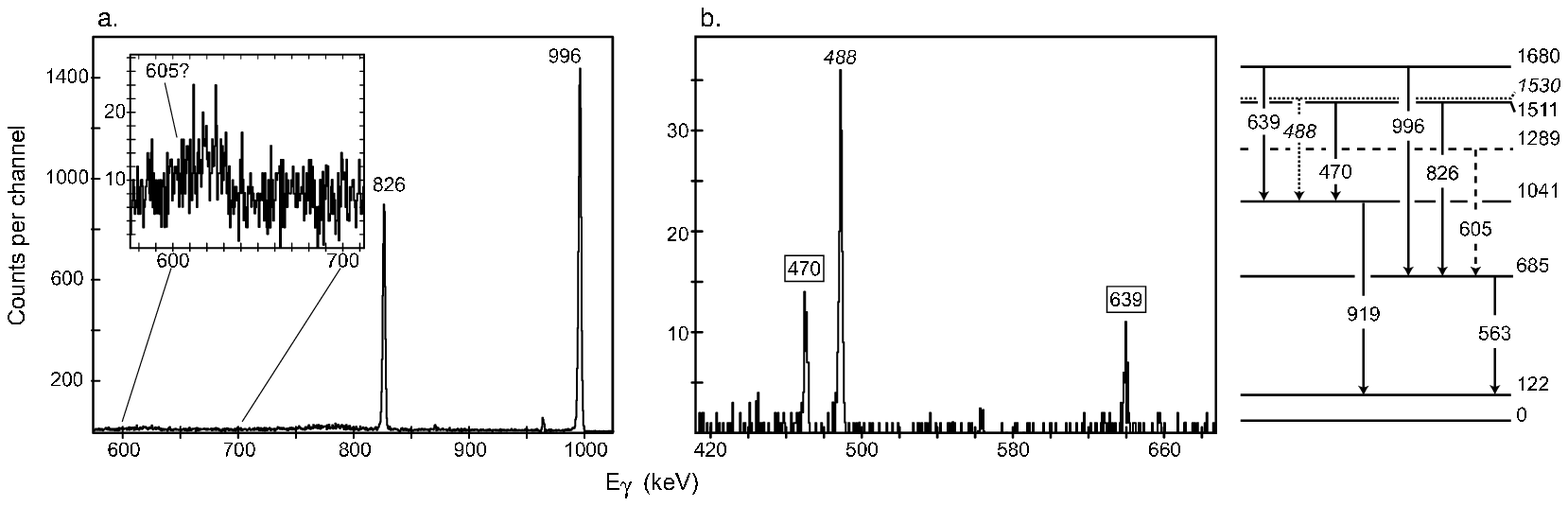}
\end{figure*}
    
\paragraph*{}
  The present study refutes the evidence of a level at 1289.94~keV.
  This level is adopted in the {\em Nuclear Data Sheets} \cite{Artna-Cohen1996} on the basis of a personal communication.
  It was established with transitions of 1290.0, 1168.16, and 605.0~keV feeding the ground state, 121.78, and 684.70~keV levels, respectively.
  Figure \ref{isomer decay}a shows the energy region where the 605.0~keV $\gamma$-ray line would be observed in the 563~keV (685 $\rightarrow$ 122) $\gamma$-ray gated coincidence spectrum.
  We set an upper limit of $<$0.0014.
  We set intensity limits of $<$0.0022 and $<$0.0011, respectively, for the 1168 and 1290~keV lines.
  The corresponding intensities given in the {\em Nuclear Data Sheets} are, $E_\gamma$($I_\gamma$):  1290 (0.0063), 1168 (0.042), 605 (0.028); of these transitions, the largest expected $N_{12}$ product would be from the 605~keV transition in the 563~keV gate, which is shown in Fig. \ref{isomer decay}a.

\paragraph*{}
  Evidence for some of the new transitions assigned to the $\eum \rightarrow \sm$ decay scheme are presented in Fig. \ref{isomer decay}.
  The two new transitions at 470 and 639~keV in coincidence with the 919 keV transition $3^-_1 \rightarrow 2^+_1$, shown in Fig. \ref{isomer decay}b, establish the population of the $3^-_1$ level in the decay of the $J^{\pi}=0^-$ parent isotope, $\eum$.
  The 470~keV $\gamma$ ray is from the 1511 $1^-_2$ level.
  The 639~keV $\gamma$ ray is from the 1680 $1^-_3$ level.

\section{DISCUSSION OF EXPERIMENTAL RESULTS}

\subsection{Decay of $\eug$}\label{eugdecay}

\paragraph*{}
  The decay of $\eug$ ($T_{1/2} = 13.6$ year) has been the subject of many experimental studies with a number of stated goals.
  This decay has been established as a leading secondary standard for energy and photopeak efficiency calibration of Ge detectors.
  An international coordinated comparison of $\gamma$-ray relative intensities of strong lines in the decay was completed in 1979 \cite{Debertin1979}.
  This focus on $\eug$ has continued with updated evaluations in 1991 \cite{Bambynek1991} and 2004 \cite{Vanin2004}.
  The nuclear structure interest in $\sm$ has also motivated many studies of the $\eug \rightarrow \sm$ decay scheme.
  This is taken up in Section \ref{discussion}.
    Recent experimental studies of the $\eug \rightarrow \sm$ decay scheme, which focus on issues of structure, can be found in \cite{Casten1998, Zamfir1999a, Kulp2005a}.
  Studies of $\eug$ decay have also been motivated by its ready availability as a radioactive source.
  This has led to searches for weak $\gamma$-decay branches \cite{Castro2004b} and to tests of detector systems such as for $\gamma-\gamma$ angular correlation measurements (see, e.g. \cite{Asai1997}).
    
\paragraph*{}    
  The present work focuses on weak $\gamma$-ray branches.
  A number of weak (and very weak) low-energy $\gamma$ rays are crucial to an understanding of the collective structure of $\sm$, particularly the 126, 212, and 275~keV lines.
  To obtain sufficient precision for the intensities of these lines (in order that the deduced $B(E2)$ values can resolve issues of collectivity in $\sm$), it is necessary to obtain coincidence-gated spectra and to understand the quantitative aspects of $\gamma$-line intensities extracted from such data; and further, to inspect the decay scheme being studied to ascertain where care must be taken to avoid potential sources of error.

\begin{table}[htb]
\caption{\label{weak gamma comparison} A comparison of critical $\gamma$-ray intensities in the decay of $\eug \rightarrow \sm$ from this work ($8\pi$) with reported values from {\em Nuclear Data Sheets} \cite{Artna-Cohen1996}, a recent evaluation of the $\eug$ decay scheme \cite{Vanin2004}, and a recent study on the nuclear structure of $\sm$ \cite{Zamfir1999a}.  Intensities are normalized using $I_{\gamma}(344) \equiv 100.0$ (in the case of \cite{Vanin2004}, $I_{\gamma}(1408.0) \equiv 78.5$ is assumed with this normalization).}
\begin{ruledtabular}
\begin{tabular}{D{.}{.}{1}D{.}{.}{8}D{.}{.}{8}D{.}{.}{8}D{.}{.}{8}} \\
 & \multicolumn{4}{c}{$I_{\gamma}$}\\
E_{\gamma} &  \multicolumn{1}{c}{$8\pi$} & \multicolumn{1}{c}{\cite{Artna-Cohen1996}} & \multicolumn{1}{c}{\cite{Vanin2004}} &  \multicolumn{1}{c}{\cite{Zamfir1999a}} \\
\hline
125.7 & 0.0193\,(7)   & 0.06\,(2)        & 0.071\,(24)\footnotemark[1]     &  0.0131\,(33) \\
212.6 & 0.0814\,(27) & 0.0741\,(20) & 0.0738\,(24)\footnotemark[2]  &  0.060\,(6) \\
275.5 & 0.137\,(5)     & 0.125\,(8)      & 0.122\,(6)\footnotemark[3]       &  0.31\,(4)
\end{tabular}
\end{ruledtabular}
\footnotetext[1]{Based upon two measurements:  $0.045\,(7)$ \cite{Meyer1990}, $0.091\,(10)$ \cite{Stewart1990}.}
\footnotetext[2]{From 7 measurements (0.059 - 0.091).}
\footnotetext[3]{From 7 measurements (0.098 - 0.171).}
\end{table}  
     
\paragraph*{}
  Information on the intensities of the 126, 212, and 275~keV lines, from work conducted prior to the study reported in Ref. \cite{Zamfir1999a} and the present investigation, is summarized in Table \ref{weak gamma comparison}.
  
\paragraph*{}
  The adopted \cite{Artna-Cohen1996} intensity and the recently re-evaluated \cite{Vanin2004} intensity for the 126~keV line are both based on just two measurements, $I_{\gamma}(126) = 0.045\,(7)$ \cite{Meyer1990}, $0.091\,(10)$ \cite{Stewart1990}.
  These two measurements were singles measurements and necessitated the resolution of the 126~keV line from the 122~keV line which is $\sim$1800$\times$  stronger.
  In the 563~keV $(0^+_2 \rightarrow 2^+_1)$ gate used in this work, the 122~keV line is $\sim$25$\times$ stronger than the 126~keV line (and is due predominantly to coincidences from the 563.96 and 566.41~keV lines in the gate).
  Nevertheless, the 126~keV line is well resolved in the 563~keV gate, cf. Fig \ref{key data}b, has a peak:background ratio of $\sim$15:1, and a fit to its area yields an uncertainty of $\pm 1.5\%$.
  In contrast, a peak:background ratio of $\sim$1.5:1 was achieved when gating from above in \cite{Zamfir1999a} (cf. Fig. 3c in \cite{Zamfir1999a}) and an uncertainty of $\pm 25\%$ was reported.  
  (We reiterate that gating from above to determine the intensity of the 126~keV line incurs an attenuation factor of $\sim$7, even using the summed gates employed in \cite{Zamfir1999a}.)
  Our intensity for the 126~keV line is $47\%$ larger than that of Zamfir, et al. \cite{Zamfir1999a}, and therefore our $B(E2)$ for this key collective transition is $47\%$ larger.
        
\paragraph*{}    
  The adopted \cite{Artna-Cohen1996} and re-evaluated \cite{Vanin2004} intensities for the 212~keV line are based on seven measurements in the range 0.059 - 0.091.
  The result from \cite{Zamfir1999a} is at the lower end of this range.
  In \cite{Zamfir1999a}, coincidence gating from above was used (cf. Fig. 3d in \cite{Zamfir1999a}) which incurs an attenuation by a factor of 6.
  (There is a further impediment to the gating method used in \cite{Zamfir1999a}; gating from above evidently gives strong Compton artifacts in the spectra which may be distorting the real peaks.)
  Our intensity for the 212~keV line is $36\%$ larger than that of \cite{Zamfir1999a}, and therefore our $B(E2)$ for the $4^+_2 \rightarrow 2^+_2$ transition, also a key collective transition, is $36\%$ larger.
  
\paragraph*{}    
  The adopted \cite{Artna-Cohen1996} and re-evaluated \cite{Vanin2004} intensities for the 275~keV line also are based on seven measurements, with values which are in the range 0.098 - 0.171.
  The result from \cite{Zamfir1999a} is nearly twice the highest reported value and is $\sim$30 standard deviations from the re-evaluated intensity \cite{Vanin2004}.
  This difference in the 275~keV $\gamma$-ray intensity was emphasized in \cite{Zamfir1999a}  as a key result because it changed the multipolarity of the transition from non-collective pure $M1$ to collective pure $E2$.
  However, the $I_{\gamma}(275)$ reported in \cite{Zamfir1999a}  is probably erroneous because of double counting of its intensity when observed in coincidence.
  This can be understood from Fig. \ref{unresolved doublets}.
  In \cite{Zamfir1999a}, the intensity of the 275~keV line was deduced from summed-coincidence gates on the 444, 494, 644, and 671~keV feeding transitions (cf. Fig. 3b in \cite{Zamfir1999a}).
  This results in severe Compton artifacts and double counting of 275~keV coincidences through the 444~keV gate contributions (note in Fig. \ref{unresolved doublets} that the 275~keV transition is both preceded {\em and} followed by $\sim$444~keV transitions).
  This double-counting is simply corrected using Eq. \ref{Ndef} and we arrive at a corrected intensity for the 275~keV line, as observed in \cite{Zamfir1999a}, of $0.15\,(2)$.
  This corrected value is in agreement with our value of $0.137\,(5)$ and is consistent with the adopted and re-evaluated intensities of $0.125\,(8)$ \cite{Artna-Cohen1996} and $0.122\,(6)$ \cite{Vanin2004}, respectively.
  Thus, we conclude that this correction was not made in \cite{Zamfir1999a}.
  The outcome of this is that the 275~keV $\gamma$-ray intensity, when combined with the conversion electron intensities of Goswamy et al. \cite{Goswamy1991} yields $\alpha_K(275) = 0.097\,(8)$ (expt.) cf.  $\alpha_K(275,M1) = 0.0862\,(12)$ (theory)  \cite{Kibedi2005}.
  Therefore, the 275~keV transition is (nearly) pure $M1$ and is non-collective.

\paragraph*{}  
  In the present work we also observe the 401~keV $2^+_3$ (1086) $\rightarrow 0^+_2$ (685) transition.
  The recent study \cite{Zamfir1999a} that sought to measure an intensity for this transition by gating from above was unsuccessful in observing this transition.
  The reason is simply understood in terms of coincidence gating sensitivity from above and below.
  Thus, gating from above using 444+494+644+671~keV gates (cf. Fig. 2b in \cite{Zamfir1999a}), ($\sum I_{\gamma})B_{\gamma}(401) = 11.2\times 2.7\times 10^{-5} = 3.1\times 10^{-4}$, whereas gating from below using the 563~keV transition (cf. Fig. \ref{key data}b), $I_{\gamma}(401)B_{\gamma}(563) = 0.00255 \times 0.97 = 2.4\times 10^{-3}$, i.e., gating from below is $8\times$ more sensitive.
  We note that in \cite{Zamfir1999a}, $I_{\gamma}(401) < 0.008$ was obtained:  a gate from below therefore would have been sensitive at the 0.001 intensity level, sufficient to observe the line.
  (The recent study of Castro et al. \cite{Castro2004b} observes the 401~keV line with a reported intensity per $10^5$ decays of $\eug$, of $0.59\pm 0.06$ which can be compared with our result, per $10^5$ decays of $\eug$, of $0.68\pm 0.06$.)
  We concur with the point made in \cite{Zamfir1999a} that the $B(E2; 401)$, which we determine in the present work to be 0.035 W.u., is remarkably weak.
  However, we discuss a different interpretation to that of \cite{Zamfir1999a} in Sect. \ref{discussion}.

\paragraph*{}  
  This work augments the $\eug \rightarrow \sm$ decay scheme with 42 new transitions, an increase of $50\%$.
  In the late stages of this work we became acquainted with a metrological study of this decay by Castro et al. \cite{Castro2004b}.
  They arrive at essentially the same number of additions and, thus, agree with the basic decay scheme aspects of the present work in a metrological context.
  We note that Castro et al. used a running time of 3 months ($\sim$2000 hours); which can be compared with a running time of 600 hours in \cite{Zamfir1999a} and 252 hours in this work.
  Castro et al. used four (15-40$\%$) detectors \cite{Castro2004b}, whereas in \cite{Zamfir1999a} 21 detectors, including three segmented-clover detectors, one $70\%$ detector, 16 $\sim$25$\%$ detectors, and one LEPS detector were used.
  In \cite{Zamfir1999a}, the source strength was 7$\mu$Ci, in \cite{Castro2004b} 27$\mu$Ci, whereas we used 50$\mu$Ci;  source-to-detector distances were comparable between \cite{Zamfir1999a} and this work.
  Castro et al. used an innovative coincidence analysis technique to extract intensities of weak $\gamma$ rays, which involved two-dimensional fitting to the $\gamma-\gamma$ coincidence matrix.
  Comparable numbers of coincidence events appear to have been obtained in all three experiments (Castro et al. $\sim$10$^9$ events, this work $6\times 10^8$ events).
  It appears that the main differences between the present work and Zamfir et al. \cite{Zamfir1999a} and Castro et al. \cite{Castro2004b} is in the subtraction of random coincidences, cf. Fig. 1c and Fig. 3b in \cite{Zamfir1999a}, and cf. Fig. 3 and Fig 1 in \cite{Castro2004b} (and excepting the less sensitive gating choices made in \cite{Zamfir1999a}).

\subsection{Decay of $\eum$}\label{eumdecay}

\paragraph*{}
  The decay of $\eum$ ($T_{1/2} = 9.3$ h) has been the subject of only a few experimental studies.
  All of these studies have been directed towards issues of structure in $\sm$ (and $\gd$).
  The most recent {\em Nuclear Data Sheets} incorporate results from all of the previously-reported experimental studies.
  
\paragraph*{}
  The main purpose of the present study was to investigate the completeness of the $\sm$ level scheme.
  In this respect, the most important outcome of this study is the refutation of a level at 1289.94~keV which was adopted in \cite{Artna-Cohen1996}.
  At this low an energy, it is essential to have complete level information in order to discuss collective behavior:  the adopted \cite{Artna-Cohen1996} 1289.94~keV level was assigned $J^\pi = 1,2^+$ and so could have been an important feature of low-energy collectivity in $\sm$.
  We note that its adoption was based on a single study of $\eum$ (9.3 h) which was only reported as a private communication to {\em Nuclear Data Sheets} \cite{Artna-Cohen1996}.
  
\paragraph*{}
  Compared to the decay of $\eug$, the $\eum$ decay uniquely populates the $1^-_2$ (1510.8) and $1^-_3$ (1680.1) levels;  and it more strongly populates the $0^+_2$ (684.7), $1^-_1$ (963.4) and $0^+_3$ (1082.9) levels.
  Thus, the $1^-_1 \rightarrow 2^+_2$ (152.77) is seen uniquely in this decay.
  The negative parity states in $\sm$ are outside of the scope of the present discussion;  but a preliminary communication \cite{Garrett2005}, based in part on the present decay data, has been made.

\section{DISCUSSION OF THE STRUCTURAL IMPLICATIONS OF THIS WORK}
\label{discussion}

\paragraph*{}
  The intriguing structural features of $\sm$ and the $N=90$ isotones have been reviewed by Mackintosh \cite{Mackintosh1977}.
  Increasing $B(E2)$ values in the ground-state band suggest that the intrinsic quadrupole moment stretches with rotation, i.e., $\sm$ appears to be a ``soft'' nucleus.
  The measured mean-square charge radius changes dramatically between the ground state and the $2^+_1$ state in $\sm$ \cite{Wu1969}; this large isomer shift is cited \cite{Mackintosh1977} as a direct measure of nuclear softness.
  Two-neutron transfer reaction data \cite{Hinds1965, Bjerregaard1966, McLatchie1969, McLatchie1970, Debenham1972} indicate shape isomers and mixing may be present in the $N=90$ nuclei \cite{Kulp2003a, Kulp2005a}, and Mackintosh \cite{Mackintosh1977} pointed out that the nature of the $0^+_2$ state (assumed at the time to be $\beta$-vibrational) is unclear.

\paragraph*{}
  The $N=90$ isotones ($\sm$ in particular) have been the subject of many theoretical studies because of their unique structural features and the availability of detailed experimental data.
  Recent perspectives of the theory in this region \cite{Burke2002a, Clark2003a, Caprio2005} have focused primarily on the latest calculations of excitation energies and $B(E2)$ values.
  Burke \cite{Burke2002a} specifically addresses the interpretation \cite{Casten1998, Iachello1998, Zamfir1999a, Klug2000} of coexisting rotational and vibrational phases in $\sm$, points out that such an interpretation does not explain one- and two- nucleon transfer data for excited states in $\sm$, and establishes figures of merit to quantitatively compare theories.
  Clark {\em et al.} \cite{Clark2003a} also compares theory with experiment, and include calculations from the X(5) model \cite{Iachello2001}.  
   Caprio \cite{Caprio2005} investigates the assumed separation of $\beta$ and $\gamma$ degrees of freedom in the X(5) model and finds that these degrees of freedom are strongly coupled and cannot be treated as separable.
  
\paragraph*{}
  A thorough survey of the theory literature for $\sm$ shows that at least nine different collective models have been used to calculate extensive sets of level energies and $B(E2)$ values.    
  In the following sections, results from the asymmetric rotor model (ARM) \cite{Gupta1982}, rotational-vibrational model (RVM) \cite{Bhardwaj1983}, pairing-plus-quadruople model (PPQ) \cite{Kumar1971, Kumar1974, Gupta1983}, boson expansion theory (BET) \cite{Kishimoto1976, Tamura1979}, interacting boson model (IBM) \cite{Scholten1978, Suhonen1985, Zamfir1999a}, X(5) model \cite{Iachello2001, Bijker2003, Caprio2005}, extended phonon model (EPM) \cite{Suhonen1985}, generalized collective model (GCM) \cite{Zhang1999}, and dynamic collective model (DCM) \cite{Mitroshin2005} are compared with experimental data.
  Level excitation energies for the four lowest-lying positive-parity bands are compared with theoretical calculations in Section \ref{Ex}.
  New absolute $B(E2)$ values are calculated using the $\gamma$-ray intensities reported in Tables \ref{longlived} and \ref{shortlived} and compared with theory in Section \ref{be2s}.

\paragraph*{}  
  Burke provided an overview of the one- and two-nucleon transfer reaction data and applicable theory in his review \cite{Burke2002a}.
  A recently-published paper \cite{Kulp2005a} uses two-neutron transfer data and some of the results of this work to interpret the nature of the band built upon the $0^+_3$ state at 1083~keV.
  This result is briefly reviewed in Sect. \ref{transfer}. 
  
\paragraph*{}  
  Comparison of the experimental isomer shift and electric monopole ($E0$) transition strength, $\rho^2(E0)$, data with theoretical calculations is made in Section \ref{e0s}.
  The isomer shift has been calculated using the cranking model \cite{Marshalek1968, Meyer1973}, PPQ \cite{Kumar1971, Kumar1974, Gupta1983}, BET \cite{Tamura1979}, IBM \cite{Scholten1978}, and IBM-2 \cite{Janecke1984} models.
  Kumar \cite{Kumar1974} has predicted many $\rho^2(E0)$ values using the pairing-plus quadrupole model.
  Additional PPQ \cite{Gupta1983} and IBM \cite{Scholten1978, Passoja1986} $\rho^2(E0)$ calculations are included for comparison. 
    
\paragraph*{}
  General conclusions based upon the results of this work (including the comparisons in Sect. \ref{Ex} - \ref{e0s}) are presented in Sect. \ref{remarks}.
  
\subsection{\label{Ex} Level excitation energies}

\begin{figure*}[htb]
\caption{\label{energies}  (Color online)  Experimental values for level excitations in the first four bands of $\sm$ plotted as a function of $J(J+1)$.
  Level excitations calculated for collective models are plotted for comparison.
  Theoretical values are denoted by letters A-Q (cf. legend above the figure), indicating order of decreasing energy of the $4^+_1$ state.}
\includegraphics[width=17.8 cm]{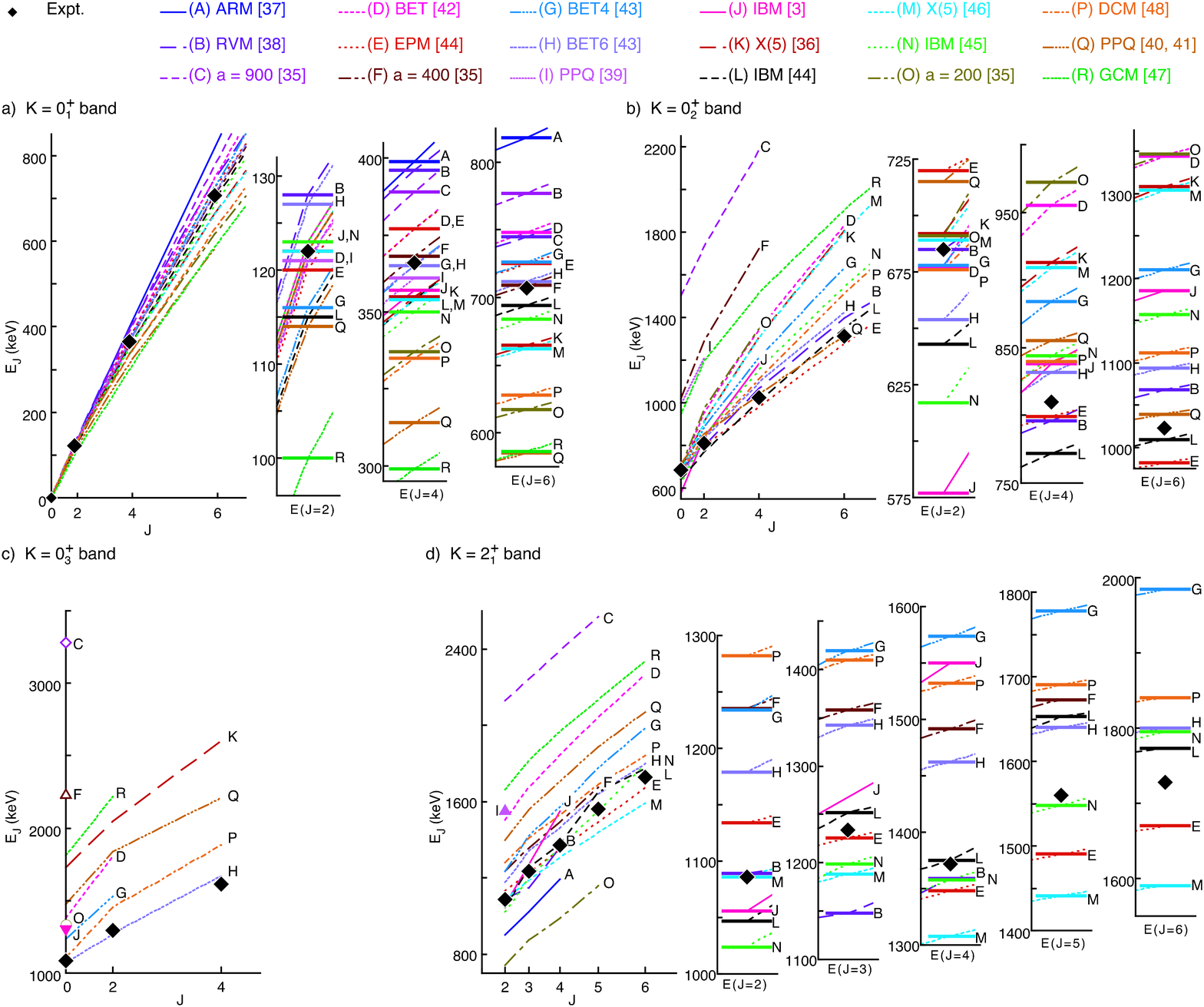}
\end{figure*}

\paragraph*{}
  Experimental level energies for the first four positive-parity bands in $\sm$ are plotted versus $J(J+1)$ in  Fig.~\ref{energies}.
  Panels a-d show, respectively, the $K^\pi = 0^+_1$ ground-state band up to spin $J^\pi=6^+$, the $K^\pi = 0^+_2$ band to $J=6^+$, the $K^\pi = 0^+_3$ band to $J=4^+$, and the $K^\pi = 2^+_1$ band to $J=6^+$.
  Magnified views of the Individual levels in the $K^\pi = 0^+_1$, $K^\pi = 0^+_2$, and $K^\pi = 2^+_1$ bands are shown next to the band plots for clarity.
 
\paragraph*{}  
  All of the models considered correspond reasonably well with experiment in the ground-state band, as shown in Fig. \ref{energies} a.
  The energy of the $J=4$ state has been used as reference for labeling purposes, i.e., (A) the asymmetric rotor model \cite{Gupta1982} has the highest calculated energy, while (R) the geometric collective model \cite{Zhang1999} has the lowest calculated energy.
  The ARM (A) \cite{Gupta1982}, X(5) (C, F, K, M, O) \cite{Iachello2001, Bijker2003, Caprio2005}, and DCM (P) \cite{Mitroshin2005} models, which have apparently used the the energy of the $2^+_1$ state for a scale factor, are not marked individually for this level.
  Differences in model moments of inertia are apparent in the $\pm 100$~keV spread of values around the experimental data at $J=6$.
  
\paragraph*{}  
  Differences in the intrinsic excitations in each model, shown in Fig. \ref{energies} b, are reflected in the very different bandhead energies and moments of inertia.
  The excitation energies predicted by the X(5), $a=900$ (C) \cite{Caprio2005}, EPM (E) \cite{Suhonen1985}, PPQ (I) \cite{Gupta1983}, and GCM (R) \cite{Zhang1999} calculations disagree with experiment such that these values are not included in the magnified views of the $J = 2, 4, 6$ level energies.
  The steeper slope of the plotted IBM (J) \cite{Zamfir1999a} values, compared with the experimental data, indicates that the rotational moment of inertia implied in this model is smaller than that realized by this band. 
  
\paragraph*{}  
  The few models which have published calculated energies for the third band in $\sm$, the $K^\pi = 0^+_3$ band at 1082 keV, are shown in Fig. \ref{energies} c.
  The X(5) $a = 900, 400, 200$ (C, F, O, respectively )\cite{Caprio2005} and IBM (J) \cite{Zamfir1999a} calculations only include the band-head energy for this band.
  The values predicted by the X(5) $a = 900, 400$ (C, F, respectively) \cite{Caprio2005}, X(5) (K) \cite{Iachello2001}, and GCM (R) \cite{Zhang1999} disagree with the experimental band-head energy by more than 1~MeV and have a considerably poorer fit to data than the other models.
  Only the sixth-order BET (H) \cite{Tamura1979} calculations appear to have good agreement with the level excitation energies for this band.

\paragraph*{}  
  Data and theoretical values for the excitation energies of the levels in the $K^\pi = 2^+_1$ ``gamma'' band are shown in Fig. \ref{energies} d.
  Calculated energies for the X(5) $a =900, 400$ (C,F)\cite{Caprio2005}, BET (D)\cite{Kishimoto1976} 
EPM (E)\cite{Suhonen1985}, PPQ (I)\cite{Gupta1983}, PPQ (Q)\cite{Kumar1971, Kumar1974}, and GCM (R)\cite{Zhang1999} models are very high compared with experiment, while the values calculated for the ARM (A)\cite{Gupta1982} and PPQ (Q)\cite{Kumar1971, Kumar1974} are low;  these outliers are not plotted in the magnified views of the $J = 2, 3, 4, 5, 6$ level energies.
    The smaller moment of inertia for the calculated  IBM (J)\cite{Zamfir1999a} values again result in a steeper slope for this band (cf. Fig.\ref{energies} b). 
      
\subsection{\label{be2s} $\mathbf{B(E2)}$ values}

\paragraph*{}
  New branching fractions for transitions in $\sm$ can be computed using the $\gamma$-ray intensities reported in Tables \ref{longlived} and \ref{shortlived}.
  These results, when combined with internal conversion data and level lifetime measurements tabulated in {\em Nuclear Data Sheets} \cite{Artna-Cohen1996} and new lifetime measurements reported in \cite{Klug2000} and \cite{Zamfir2002b} provide absolute $B(E2)$ values.
  The experimental absolute $B(E2)$ values calculated in this way and expressed as ratios with $B(E2; 2^+_1 \rightarrow 0^+_1)$ are presented in Table \ref{be2 comparison}.

\begin{turnpage}
\begingroup
\squeezetable
\begin{table}[H] 
\caption{\label{be2 comparison} Selected experimental $B(E2)$ values from this work and {\em Nuclear Data Sheets} \cite{Artna-Cohen1996}, expressed as a ratio with $B(E2; 2^+_1 \rightarrow 0^+_1)$, compared to theoretical values.  
  Boldface numbers indicate agreement within experimental uncertainties.
  The entry ``agreement'' in the last line of the table is the total of boldface entries in each column.}
\begin{ruledtabular}
\begin{tabular}{ccccccccccccccccccccccccc} \\
Transition & Expt &  ARM & RVM  & $a=900$  & BET & EPM & $a=400$ & BET4 & BET6 & PPQ  & IBM  & X(5) & IBM  & X(5) & IBM & $a=200$ & DCM & PPQ  & GCM  \\ 
 & & (A)\cite{Gupta1982} & (B)\cite{Bhardwaj1983} & (C)\cite{Caprio2005} & (D)\cite{Kishimoto1976} & (E)\cite{Suhonen1985} & (F)\cite{Caprio2005} & (G)\cite{Tamura1979} & (H)\cite{Tamura1979} & (I)\cite{Gupta1983} & (J)\cite{Zamfir1999a} & (K)\cite{Iachello2001} & (L)\cite{Suhonen1985} & (M)\cite{Bijker2003} & (N)\cite{Scholten1978} & (O)\cite{Caprio2005} & (P)\cite{Mitroshin2005} & (Q)\cite{Kumar1971, Kumar1974} & (R)\cite{Zhang1999} \\
\hline
$2_1^+ \rightarrow 0_1^+$ & 1  & 1 & 1 & 1 & 1 & 1 & 1 & 1 & 1 & 1 & 1 & 1 & 1 &  & 1 & 1 & 1 & 1 & 1\\ 
$4_1^+ \rightarrow 2_1^+$ & 1.45\,(4) & \bf{1.45} & 1.58 & \bf{1.48} & 1.52 & \bf{1.49} & 1.53 & \bf{1.47} & \bf{1.46} & 1.52 & 1.50 & 1.58 & \bf{1.41} &  & \bf{1.45} & 1.60 & \bf{1.41} & 1.53 & 1.51\\ 
$6_1^+ \rightarrow 4_1^+$ & 1.70\,(5) & 1.64 &  & \bf{1.72} & 1.76 & 1.76 & 1.83 & \bf{1.66} & 1.62 &  &  & 1.98 & 1.51 &  &  & 1.97 & 1.56 & 1.83 & \\ 
$8_1^+ \rightarrow 6_1^+$ & 1.98\,(11) & 1.65 &  & \bf{1.90} & \bf{1.95} & 2.10 & \bf{2.07} & 1.74 & 1.65 &  &  & 2.27 & 1.50 &  &  & 2.27 & 1.66 &  & \\ 
$10_1^+ \rightarrow 8_1^+$ & 2.22\,(21) & 1.69 &  & \bf{2.06} &  & 2.51 & \bf{2.27} &  &  &  &  & 2.61 & 1.43 &  &  & 2.53 &  &  & \\[0.1 in] 
$0_2^+ \rightarrow 2_1^+$ & 0.229\,(28) &  & 0.469 & \bf{0.210} & 0.337 & 0.134 & 0.325 & \bf{0.232} & 0.178 & 0.266 & 0.382 & 0.630 & 0.034 &  &  & 0.510 & 0.060 & 0.318 & 0.319\\[0.1 in] 
$2_2^+ \rightarrow 0_2^+$ & 1.16\,(11) &  &  & 0.70 & 0.61 & 1.59 & 0.66 & 0.67 & 0.67 & 1.03 & 0.62 & 0.79 & 0.66 &  &  & 0.71 & 0.19 & \bf{1.15} & 0.67\\ 
$2_2^+ \rightarrow 0_1^+$ & 0.0067\,(6) &  & 0.028 & 0.017 & 0.013 & 0.002 & 0.041 & 0.007 & 0.010 & 0.004 & 0.001 & 0.020 & 0.008 &  & 0.003 & 0.021 & 0.003 & 0.005 & 0.022\\ 
$2_2^+ \rightarrow 2_1^+$ & 0.040\,(4) &  & 0.093 & \bf{0.042} & 0.057 & 0.019 & 0.011 & 0.046 & \bf{0.037} & 0.046 & 0.069 & 0.090 & 0.006 &  & \bf{0.041} & 0.035 & 0.007 & \bf{0.044} & 0.065\\ 
$2_2^+ \rightarrow 4_1^+$ & 0.125\,(14) &  & 0.422 & \bf{0.128} & 0.178 & 0.179 & 0.200 & 0.074 & 0.104 & 0.143 & \bf{0.139} & 0.360 & 0.022 &  & 0.093 & 0.263 & 0.034 & 0.212 & 0.181\\[0.1 in] 
$4_2^+ \rightarrow 2_2^+$ & 1.8\,(3) &  &  & 1.0 & 1.1 & 2.256 & 1.1 & 1.1 & 0.9 & \bf{1.8} & 1.0 & 1.2 & 0.8059 &  &  & 1.2 & 1.0 & \bf{1.7} & 1.2\\ 
$4_2^+ \rightarrow 2_1^+$ & 0.0052\,(9) &  & 0.0042 & 0.0110 & 0.0109 & 0.0007 & 0.0160 & \bf{0.0059} & 0.0074 & 0.0010 & 0.0007 & 0.0100 & 0.0104 &  &  & 0.0130 & 0.0014 & 0.0003 & 0.0145\\ 
$4_2^+ \rightarrow 4_1^+$ & 0.042\,(8) &  & 0.104 & \bf{0.047} & 0.052 & 0.029 & 0.019 & \bf{0.039} & 0.024 & \bf{0.042} & 0.056 & 0.060 & 0.003 &  &  & 0.020 & 0.016 & \bf{0.041} & 0.065\\ 
$4_2^+ \rightarrow 6_1^+$ & 0.11\,(3) &  &  &\bf{ 0.11} & 0.15 &  & 0.17 & \bf{0.10} & 0.07 & 0 &  & 0.28 &  &  &  & 0.21 & 0.06 & 0.22 & \\[0.1 in] 
$2_3^+ \rightarrow 0_1^+$ & 0.0258\,(18) & 0.0521 & 0.0224 & 0.0450 & 0.0387 & 0.0397 & 0.0400 & 0.0728 & 0.0743 & 0.0369 & 0.0208 &  & \bf{0.026} &  & 0.0152 & 0.0710 & 0.0514 & 0.0338 & 0.0217\\ 
$2_3^+ \rightarrow 2_1^+$ & 0.065\,(4) & 0.140 & 0.046 & 0.078 & 0.040 & 0.090 & 0.187 & 0.074 & 0.079 & 0.077 & 0.021 &  & 0.129 & 0.039 & 0.015 & 0.299 & 0.096 & 0.079 & 0.029\\ 
$2_3^+ \rightarrow 4_1^+$ & 0.0049\,(4) & 0.0173 & 0.0103 & 0.007 & 0.0077 & 0.0113 & 0.0000 & 0.0104 & 0.0089 & 0.0096 & 0.0278 &  & 0.0074 & 0.0020 & 0.0114 & 0.005 & 0.0143 & 0.0040 & 0.0072\\ 
$2_3^+ \rightarrow 0_2^+$ & 0.00025\,(3) &  &  & 0.00400 & 0.00230 & 0.01471 & 0.03900 & 0.02694 & 0.03120 &  & 0.01389 &  & 0.00735 & 0.00230 &  & 0.00300 & 0.01029 & 0.00119 & 0.00036\\ 
$2_3^+ \rightarrow 2_2^+$ & 0.087\,(6) &  & 0.012 & 0.00000 & 0.113 & 0.007 & 0.187 & 0.275 & 0.208 &  & 0.639 &  & 0.042 & 0.001 &  & 0.299 & 0.010 & 0.076 & \bf{0.087}\\
Agreement & & 1 & 0 & 9 & 1 & 1 & 2& 6 & 2 & 2 & 1 & 0 & 2 & 0 & 2 & 0 & 1 & 4 & 1 
\end{tabular}
\end{ruledtabular}
\end{table} 
\endgroup
\end{turnpage}

\paragraph*{}
  In addition to experimental $B(E2)$ values, Table \ref{be2 comparison} lists theoretical values for comparison.
  Agreement within experimental error between experimental data and theoretical values is indicated by boldface print in Table \ref{be2 comparison}. 
  A tally of agreement is listed for each theory, with the most successful calculations provided by the X(5) $a = 900$ (C) \cite{Caprio2005} (9 values agree with experimental data), BET (D) \cite{Kishimoto1976} (6 values in agreement with data), and PPQ (I) \cite{Gupta1983} (4 values agree with data).

\begin{figure*}[htb]
\caption{\label{theoretical be2s}  (Color online)  Theoretical $B(E2)$ values expressed as a ratio of the $B(E2; 2_1^+ \rightarrow 0_1^+)$ value compared with experimental values.  The experimental value is marked by a white line on a gray background, representing the mean value and experimental uncertainty, respectively.  The legend provides a key to the theoretical values.  N.B. the $2^+_3 \rightarrow 0^+_2$ 401~keV and $2^+_3 \rightarrow 2^+_2$ 275~keV transitions are plotted using  logarithmic scales.}
\includegraphics[width=17.8 cm]{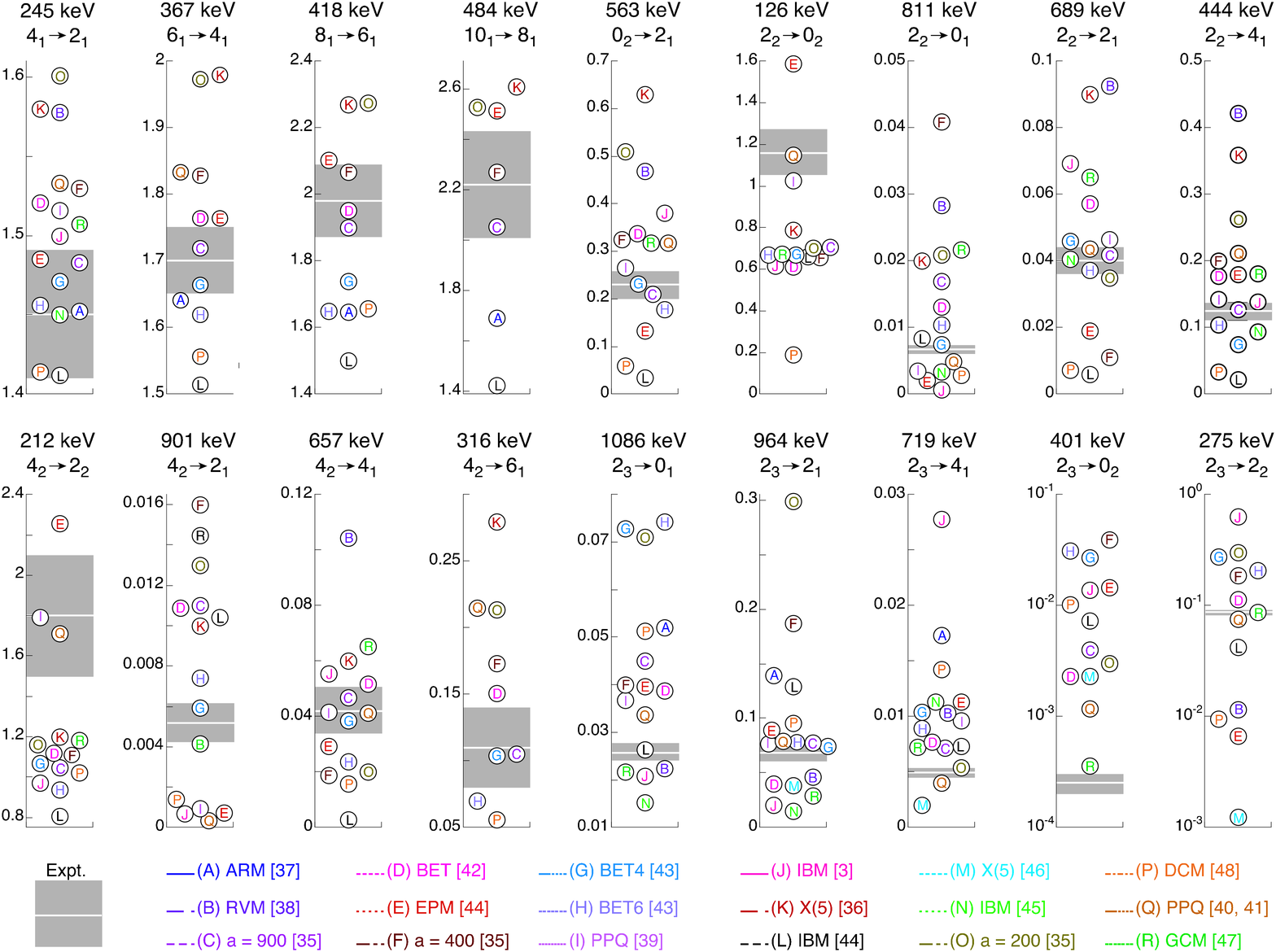}
\end{figure*}

\paragraph*{}
  While a useful ``scorecard,'' the comparison presented in Table \ref{be2 comparison} does not provide insight into which $\gamma$-ray transitions are key to understanding the structure of $\sm$.
  Figure \ref{theoretical be2s} shows the individual transitions, again expressed as ratios with $B(E2; 2^+_1 \rightarrow 0^+_1)$.
  Experimental data are plotted as a white line on a gray background, which represents the uncertainty in the measured value.
  Calculated theoretical values are plotted by reference letter (A-Q) normalized to the energy of the $J=4$ state (cf. Sect. \ref{Ex}).
  
\paragraph*{}  
  The weak $\gamma$-ray transitions that play a crucial role in understanding the low-energy collective structure of $\sm$ become apparent through the comparison in Fig. \ref{theoretical be2s}.
  In particular, the 126 keV ($2^+_2 \rightarrow 0^+_2$), 212 keV ($4^+_2 \rightarrow 4^+_2$), 275 keV ($2^+_3 \rightarrow 2^+_2$), and 401 keV ($2^+_3 \rightarrow 0^+_2$) $\gamma$-ray transitions emerge as key comparison points.
  While other transitions in Fig. \ref{theoretical be2s} have a few calculated values which agree with experiment and a modest, uniform spread of values which do not agree, the situation is different for the 126, 212, 275, and 401 keV transitions, as discussed below.
  
\paragraph*{}
  Only the PPQ (I,Q) calculations \cite{Gupta1983, Kumar1971, Kumar1974} are in agreement with the data for the $2^+_2$ (810) $\rightarrow 0^+_2$ (685) 126~keV transition.
  Most of the other theoretical values are grouped at a lesser value between $0.6-0.8$, while the experimental value, based upon the intensity measurements from this work (cf. Table \ref{longlived}) , is $1.16\,(11)$.
  This indicates that most models predict a smaller $B(E2)$ value between the $2^+$ and $0^+$ states in the excited $K^\pi = 0^+_2$ band than between the analog states in the ground-state band.
  
\paragraph*{}  
  Using the intensities listed in Table \ref{longlived} and the adopted lifetime of the $2^+_2$ level at 810~keV \cite{Artna-Cohen1996}, an absolute $B(E2; 2^+_2 \rightarrow 0^+_2) = 167\,(16)$ W.u. is determined for the 126 keV ($2^+_2 \rightarrow 0^+_2$) $\gamma$ ray transition.
  In the ground-state band, the 122 keV $\gamma$-ray transition has an absolute $B(E2; 2^+_1 \rightarrow 0^+_1) = 144\,(3)$ W.u., a smaller value.
  This indicates that the precisely measured intensity of the $2^+_2$ (810) $\rightarrow 0^+_2$ (685) 126~keV transition is one of the most important results of this work:  The larger $B(E2)$ value in the excited $K^\pi = 0^+_2$ band is characteristic of a more collective structure and could be interpreted as a coexisting shape with a larger deformation than that of the ground-state band.

\paragraph*{}
  A second result of importance is indicated by the $B(E2)$ ratio for the $4^+_2$ (1023) $\rightarrow 2^+_2$ (810) 212~keV transition (cf. Fig. \ref{theoretical be2s}); the calculated ratios again indicate a grouping of most models which differs from the experimental data.
  The measured intensity of the 212~keV transition (cf. Table \ref{longlived}) determines a $B(E2; 4^+_2 \rightarrow 2^+_2) = 255\,(45)$ W.u. when combined with the new lifetime measurement for the $4^+_2$ state \cite{Klug2000}.
   This value is greater than the $B(E2; 4^+_1 \rightarrow 2^+_1) = 209\,(3)$ W.u. for the 245 keV $\gamma$-ray transition in the ground-state band.
  This larger $B(E2)$ value again is characteristic of a more collective structure (in contrast to the conclusions of \cite{Klug2000}), providing additional evidence that the $K^\pi = 0^+_2$ band has a larger deformation than that of the ground-state band.
  
\paragraph*{}
  The $B(E2)$ ratio for the $2^+_3$ (1086) $\rightarrow 2^+_2$ (810) 275~keV $\gamma$ ray is plotted in Fig. \ref{theoretical be2s} on a logarithmic scale due to the very wide spread of calculated theoretical values.
  The BET (D)\cite{Kishimoto1976}, PPQ (Q)\cite{Kumar1971, Kumar1974}, and GCM (R)\cite{Zhang1999} calculations agree reasonably well with the experimental data;  but other calculated values differ by up to two orders of magnitude.
  The distribution of calculated theoretical values indicates that the $2^+_3 \rightarrow 2^+_2$ transition is very sensitive to the nature and interaction of $K^\pi = 0^+$ and $K^\pi = 2^+$ bands in the models considered.
  
\paragraph*{}
  By itself, the intensity of the $2^+_3$ (1086) $\rightarrow 2^+_2$ (810) 275~keV $\gamma$ ray transition is a third result of importance, as discussed in Sect. \ref{eugdecay}.
  Using the 275~keV $\gamma$-ray intensity in Table \ref{longlived} and the previously measured \cite{Goswamy1991} relative internal conversion electron intensity, the deduced multipolarity of the $2^+_3$ (1086) $\rightarrow 2^+_2$ (810) transition is consistent with pure $M1$, i.e., it lacks quadrupole collectivity.
  This result is in disagreement with that of \cite{Zamfir1999a} (cf. Fig. \ref{unresolved doublets} and the accompanying discussion of the Figure in Sect. \ref{eugdecay}), and is inconsistent with the interpretation of the $3^+_1 \rightarrow 2^+_3  \rightarrow 2^+_2$ ($148-275$ keV) cascade as a collective quadrupole 3 phonon - 2 phonon - 1 phonon sequence \cite{Zamfir2002b}.  

\paragraph*{}
  Of all the theoretical calculations, only the GCM (R)\cite{Zhang1999} ratio approaches the very small experimental value for the $2^+_3$ (1086) $\rightarrow 0^+_2$ (685) 401~keV transition.
  Similar to the case for the 275~keV $\gamma$ ray, the theoretical calculations vary greatly, and the $B(E2; 2^+_3 \rightarrow 0^+_2)/B(E2; 2^+_1 \rightarrow 0^+_1)$ ratio is displayed using a logarithmic scale in Fig. \ref{theoretical be2s}.
  Also, like the 275~keV $\gamma$ ray, the distribution of theoretical values indicates a sensitivity to the nature and interaction of $K^\pi = 0^+$ and $K^\pi = 2^+$ bands in the models considered.

\paragraph*{}  
  The 401 and 275~keV transitions both de-excite the $2^+_3$ (1086) level, which would be interpreted as the head of a $K^\pi=2^+$ band (a ``$\gamma$-vibrational'' band in the standard Bohr-Mottelson picture \cite{Bohr1975b}).
  The present work suggests that the two states fed by the 401 and 275~keV transitions constitute $0^+$ and $2^+$ members of a $K^\pi=0^+$ band at 685 and 811 keV, respectively.
  Assuming $K$ is a good quantum number for these bands and  that there is no mixing of bands, the 401 and 275~keV transitions should exhibit $B(E2)$ values approximately consistent with the Alaga rules \cite{Alaga1955}.
  This leads to an estimate for $B(E2; 2^+_3 \rightarrow 2^+_2) = 0.051$ W.u. $(=10/7\times B(E2; 2^+_3 \rightarrow 0^+_2)$), which would correspond to a partial ($E2$) decay contribution to the 275~keV transition of $0.40\%$:  This is consistent with the (near) pure $M1$ character of the 275~keV transition deduced in the present work.  

\subsection{\label{transfer} Transfer reaction data} 

\paragraph*{}
  The wealth of transfer data \cite{Hinds1965, Bjerregaard1966, McLatchie1969, McLatchie1970, Debenham1972, Nelson1973, Hirning1977, Burke2001} available for $\sm$ is reviewed by Burke \cite{Burke2002a}.
  However, the two-nucleon transfer data \cite{Hinds1965, Bjerregaard1966, McLatchie1969, McLatchie1970, Debenham1972} led to a further result based on the present work and described in detail in \cite{Kulp2005a}.
   Using a close analogy with a nearly identical $0^+_3$ state in $^{154}$Gd \cite{Kulp2003a} and a detailed consideration of one- and two-nucleon transfer data in this mass region, the $0^+_3$ state is deduced to be a pairing isomer \cite{Ragnarsson1976}.
   
\paragraph*{}
  Relative $\gamma$-ray intensities for transitions depopulating the $2^+_4$ 1293~keV and $4^+_4$ 1613~keV levels (cf. Table \ref{longlived}) show that the 210~keV and 320~keV transitions, respectively, have the strongest relative $B(E2)$ values, indicating that the $0^+_3$ (1083),  $2^+_4$ (1293), and  $4^+_4$ (1613) levels form a collective band.
  Additional $\gamma$-ray coincidence data from a multiple-step Coulomb excitation study show that a level at 2004~keV is the $6^+$ member of this band \cite{Kulp2005a}.
  The $0^+_3$ (1083) and $2^+_4$ (1293) levels are strongly populated in the $(t, p)$ stripping reaction \cite{Hinds1965, Bjerregaard1966} and very weakly populated in the $(p,t)$ pickup reaction \cite{McLatchie1969, McLatchie1970, Debenham1972}.
  This asymmetric population is a fingerprint of a pairing isomer (a state which has a smaller pairing gap than the ground state), and is suggested to be a result of the relative isolation of the ${\frac{11}{2}}^-$ [505] Nilsson configuration in the $N=90$ nuclei \cite{Peterson1984}.
  
\subsection{\label{e0s} Isomer shift and $\mathbf{E0}$ values}

\paragraph*{}
  The isomer shift, $\delta \langle R^2 \rangle / \langle R^2 \rangle$, in $\sm$ is one of the largest observed in nuclei \cite{Wu1969}.
  Using a weighted average of the measured isomer shift from muonic atom and M\"ossbauer shift  measurements, given by Wu and Willets \cite{Wu1969}, a value of $\delta \langle R^2 \rangle / \langle R^2 \rangle = 5.3\,(6) \times 10^{-4}$ is found for $\sm$.
  Isomer shift calculations for  $\sm$ have been made using the cranking model \cite{Marshalek1968, Meyer1973}, the pairing-plus quadrupole model \cite{Kumar1971, Kumar1974, Gupta1983}, boson expansion theory \cite{Tamura1979}, interacting boson model \cite{Scholten1978}, and IBM-2 \cite{Janecke1984}.
  
\begin{figure}[htb]
\caption{\label{isomer shift} Theoretical isomer shift values, expressed as a dimensionless ratio,  compared with the measured value of $\delta \langle R^2 \rangle / \langle R^2 \rangle = 5.3\,(6) \times 10^{-4}$  (where $R = 1.2 A^{1/3}$~fm).
  The white line on a gray background represents the weighted mean and experimental uncertainty,   respectively, of muonic atom and M\"{o}ssbauer measurements \cite{Wu1969}.}
\includegraphics[width=6 cm]{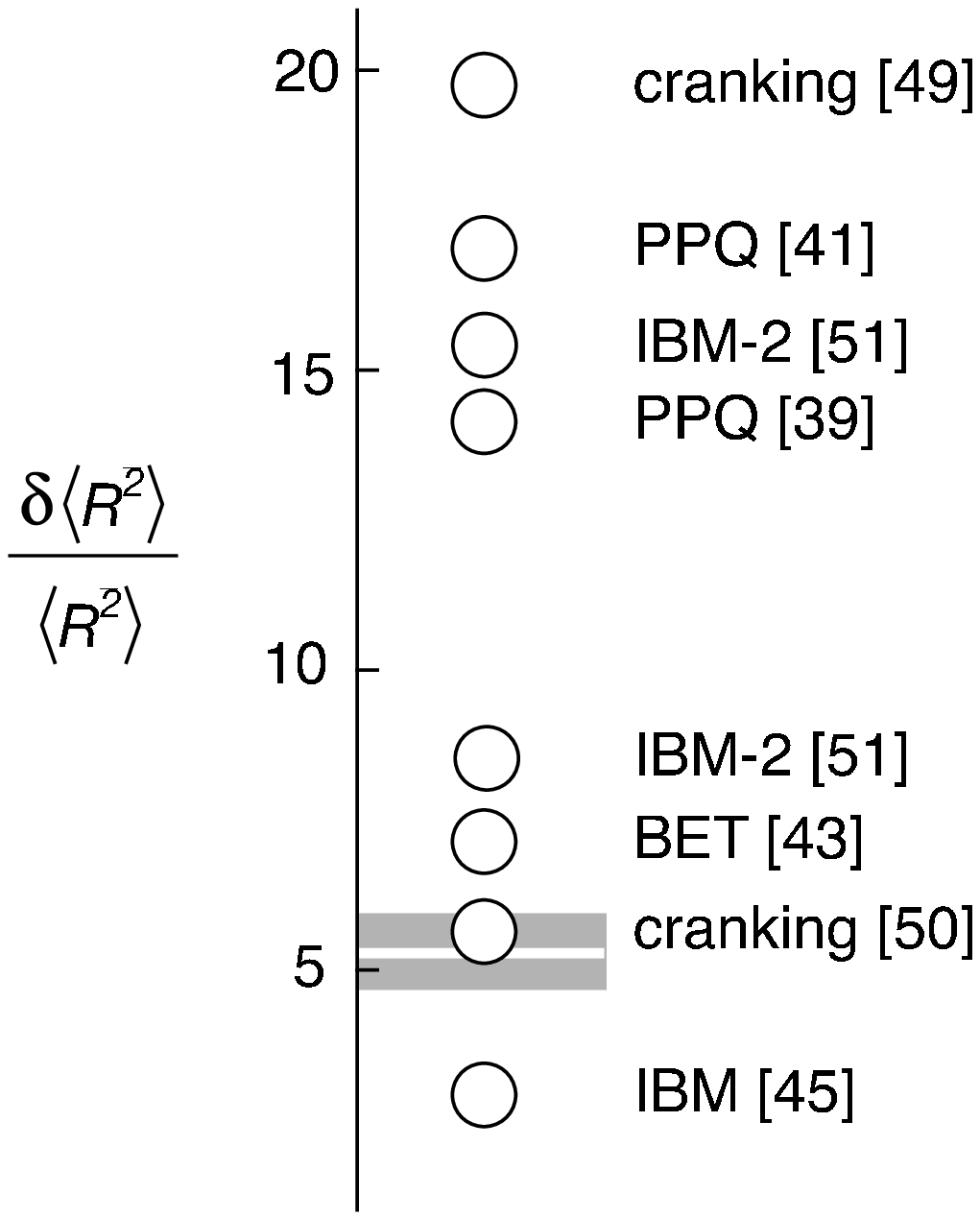}
\end{figure}

\paragraph*{}  
  Theoretical values for the isomer shift are compared with data in Fig. \ref{isomer shift}.
  Only the cranking model using a Migdal-type effective interaction \cite{Meyer1973} agrees with the experimental data.
  Meyer and Speth \cite{Meyer1973} concluded that the results of their cranking model calculations indicated a stretching effect resulted from mixing over shell model configurations with different mean-square radii near the Fermi surface in the $N=90$ region.

\paragraph*{}
  Electric monopole ($E0$) transitions are related to changes in nuclear radii in a model-independent way \cite{Wood1992, Wood1999}, and can be used to independently measure mean-square radii between configurations with the same spin, $J$, which mix.
  Table \ref{monopoles} lists the experimental values of $\rho^2(E0) \times 10^3$ for $E0$ transitions in $\sm$.
  Theoretical calculations using the pairing-plus quadrupole model \cite{Kumar1974, Gupta1983} and interacting boson model \cite{Scholten1978, Passoja1986} are included in Table \ref{monopoles} for comparison.

\begin{table}[hb]
\caption{\label{monopoles} Experimental electric monopole ($E0$) transition strengths, $\rho^2(E0) \times 10^3$, for $\sm$ \cite{Artna-Cohen1996} compared with theoretical calculations.}
\begin{ruledtabular}
\begin{tabular}{cccccc} \\
 & Expt. & DPPQ & IBM & DPPQ & IBM \\
Transition   & \cite{Artna-Cohen1996}  & \cite{Kumar1974}  &\cite{Scholten1978}  & \cite{Gupta1983}  &  \cite{Passoja1986} \\
\hline
$0^+_2 \rightarrow 0^+_1$ & 51\,(5) & 133 & 51 & 77 & 46\\
$0^+_3 \rightarrow 0^+_1$ & 0.7\,(4) & 1.44 &  &  & \\
$0^+_3 \rightarrow 0^+_2$ & 23\,(9) & 195 &  &  & \\
$\frac{0^+_3 \rightarrow 0^+_2}{ 0^+_3 \rightarrow 0^+_1}$ & 33\,(23) & 135\footnotemark[1]  &  &  & 149\\[0.2 in]
$2^+_2 \rightarrow 2^+_1$ & 72\,(7) & 132 & 40 & 104 & \\
$2^+_3 \rightarrow 2^+_1$ & 1.3\,(19) & 0.0490 & 3 & 2.30 & \\
$2^+_3 \rightarrow 2^+_2$ & 3.4\,(34) & 6.89 &  &  & \\
$2^+_4 \rightarrow 2^+_1$ &  &   3.36 &  &  & \\
$2^+_4 \rightarrow 2^+_2$ &  &   153 &  &  & \\
$2^+_4 \rightarrow 2^+_3$ &  &   2.21 &  &  & \\[0.2 in]
$4^+_2 \rightarrow 4^+_1$ & 88\,(16) & 135 & & & \\
$4^+_3 \rightarrow 4^+_1$ &    & 0.256 & & & \\
$4^+_3 \rightarrow 4^+_2$ &    & 7.23 & & & \\
$4^+_4 \rightarrow 4^+_1$ &    & 3.03 & & & \\
$4^+_4 \rightarrow 4^+_2$ &    & 116 & & & \\
$4^+_4 \rightarrow 4^+_3$ &    & 8.46 & & & \\[0.2 in]
$6^+_2 \rightarrow 6^+_1$ &    & 132 & & & \\
$6^+_3 \rightarrow 6^+_1$ &    & 0.841 & & & \\
$6^+_3 \rightarrow 6^+_2$ &    & 5.78 & & & 
\end{tabular}
\end{ruledtabular}
\footnotetext[1]{Calculated in order to compare with the ratio published by \cite{Passoja1986} .}
\end{table}  

\subsection{\label{remarks} Remarks} 

\paragraph*{}  
  Considering the comparisons presented in Sect. \ref{Ex} - \ref{e0s}, there is no single model which adequately describes all of the experimental data available for $\sm$.
  As there is no clearly superior agreement between experimental data and a single theory, we focus the remainder of the discussion only on the very recent developments of phase coexistence and critical point symmetries that are of current interest.
  
\paragraph*{} 
  The phase coexistence interpretation \cite{Iachello1998, Zamfir1999a, Klug2000, Zamfir2002b} suggested that the ground-state band of $\sm$ is a deformed rotational band, but that the $0^+_2$ level at 685~keV was the ground state of a coexisting vibrational structure.
  As discussed above (cf. Sect. \ref{be2s}), our measurements indicate that the $4^+_2 (1023) \rightarrow 2^+_2 (810)$ 212 and $2^+_2 (810) \rightarrow 0^+_2 (685)$ 126~keV $\gamma$-ray transitions in the band built upon the $0^+_2$ 685~keV level are more collective than the corresponding transitions in the ground-state band.
  Our results show that the $2^+_3 (1086) \rightarrow 2^+_2 (810)$ 275~keV transition is of nearly pure $M1$ multipolarity (cf. Sect. \ref{be2s}).
  Transfer data support the interpretation of the 1083 $0^+_3$ state as a pairing isomer \cite{Kulp2005a}, and relative $B(E2)$ values indicate that the 1293 $2^+_4$ and 1613 $4^+_4$ states are members of a rotational band built upon this structure (cf. Sect. \ref{transfer}).
  Together these results dispute the evidence of a coexisting phase interpretation of $\sm$.

\paragraph*{} 
  In contrast, the ratio of $B(E2; 4^+_2 \rightarrow 2^+_2) / B(E2; 2^+_2 \rightarrow 0^+_2)$ and very small $B(E2; 2^+_3 \rightarrow  2^+_2)$ value are in good agreement with X(5) calculations \cite{Iachello2001, Bijker2003}.
  Additionally, the $B(E2)$ values for X(5), $a =900$ (C) \cite{Caprio2005} in Table \ref{be2 comparison} have the widest agreement (within error) with experimental data.
  However, every model considered in Sect. \ref{be2s} except one agrees within error with the experimental $B(E2; 4^+_2 \rightarrow 2^+_2) / B(E2; 2^+_2 \rightarrow 0^+_2)$ ratio, all of the models predict $B(E2; 2^+_3 \rightarrow  2^+_2)/B(E2; 2^+_1 \rightarrow 0^+_1) < 10^{-3}$, and the excitation energies for the X(5), $a =900$ (C) \cite{Caprio2005} calculations are far too high in energy for the $K^\pi = 0^+_2$, $0^+_3$, and $2^+_1$ bands (cf. Fig.~\ref{energies}).

\paragraph*{}
  It should be noted that the X(5) calculations with $a = 900$ by Caprio \cite{Caprio2005} are very different from the approximate solutions \cite{Iachello2001, Casten2001a, Bijker2003, Bonatsos2004, Bonatsos2006} (which are near $a=200$ \cite{Caprio2005}).
  The X(5) interpretation is modeled using the Bohr Hamiltonian with the approximation that $u(\beta,\gamma) = u(\beta) + u(\gamma)$, taking $u(\beta)$ to be an infinite square well potential and $u(\gamma)$ to be a harmonic oscillator potential.
  This approach has the attractive feature that the $\beta$- and $\gamma$-dependent wave functions are separable and, for an infinite square well potential for $\beta$, exact analytical solutions are possible.
  The approach was necessitated at its inception \cite{Iachello2001} because, if the approximation of $\beta-\gamma$ separability had not been assumed, then the problem would have been intractable.
  The intractability of the $\beta-\gamma$ coupling problem of the Bohr Hamiltonian has now been solved by Rowe {\em et al.} \cite{Rowe2004a, Rowe2004b, Rowe2005} using its $SU(1,1) \times SO(5)$ algebraic structure.
  The $SO(5)$ part of the new approach \cite{Rowe2004a, Rowe2004b, Rowe2005}  combined with Bessel functions of irrational order has been used by Caprio \cite{Caprio2005} to obtain exact solutions of the infinite square well $\beta$ potential under a wide range of $\beta-\gamma$ coupling strengths.
    
\paragraph*{}
  Caprio \cite{Caprio2005} makes a number of key observations regarding X(5):
\begin{enumerate}
\item Even though the potential is flat in $\beta$, the $\beta-\gamma$ interaction inherent in the kinetic energy operator induces something akin to rigid $\beta$ deformation.
\item The enlarged energy spacing of the $K^\pi=0_2^+$ band is an artifact of the rigid well wall in the X(5) Hamiltonian.
\item The $\beta-\gamma$ coupling is very sensitive to the $\gamma$ rigidity.
Thus, $\gamma$ rigidity strongly influences the $K^\pi=0_2^+$ band energies and the $K^\pi=0_2^+$ to ground interband $E2$ transition strengths.
\end{enumerate}
   From these observations, we note that, while the $a=900$ solutions to X(5) provide the best agreement with the experimental $B(E2)$ values presented here, this solution does not describe a ``floppy'' nucleus.

\paragraph*{}  
  Additionally, there are features of $\sm$ which clearly are in disagreement with the entire range of X(5) solutions.
  The energy spacing of the $K^\pi=0_2^+$ band cannot be reproduced in the X(5) model.
  The experimental intra-band $E2$ strengths in the $K^\pi=0_2^+$ band are $\sim$120$\%$ of the ground band whereas in the X(5) model they are always $\sim$70$\%$.
  The $K^\pi=0_2^+$ band to ground band spin-decreasing $E2$ strengths are always at least twice the experimental values.

\paragraph*{}  
  Using the new approach to calculations within the Bohr model developed by Rowe {\em et al.} \cite{Rowe2004a, Rowe2004b, Rowe2005}, it would be possible to relax the rigid constraints on $\beta$ and further explore the model space.
  However, as noted in \cite{Caprio2005}, the $SO(5)$ centrifugal term has a particular, model-dependent set of ratios for the components of the inertia tensor (which coincidentally happen to be the same as the ratios for an irrotational flow inertia tensor).
  Due to the large $\beta-\gamma$ coupling effect of this term, as further noted in \cite{Caprio2005}, the proper $\beta$ dependence of these moments of inertia therefore needs to be determined.
  Indeed, it has been shown that the ratios of components of the inertia tensor deduced using a triaxial rotor model are uniformly at odds with irrotational values \cite{Wood2004}.

\section{SUMMARY}

\paragraph*{}
  We have studied excited states in $\sm$, populated in the decays of $\eu$, with particular attention  given to very weak $\gamma$-decay branches.
  We add many new branches to the decay of $\eug$ (13.6 year), an important secondary calibration standard.
  The present work should contribute to advancing the status of this decay for calibration purposes.
  
\paragraph*{}
  Among the very weak transitions, a number are crucial for resolving significant collective features of $\sm$.
  To this end, the technique of coincidence intensities has been developed, with attention to issues of sensitivity, possible sources of error, precision, and accuracy.
  We find and identify the cause of disagreements with the previous work \cite{Zamfir1999a}, which also depended on coincidence intensities.

\paragraph*{}
  We compare experimental level excitation energies, $B(E2)$ values, isomer shift data, and $\rho^2(E0)$ values for nine collective models.
  A new pairing isomer structure (previously reported \cite{Kulp2005a}) has been identified as a result of this work.
  No single collective model is found to agree with all of the experimental data.

\paragraph*{} 
  A valuable line of inquiry into the applicability of the Bohr Hamiltonian to $\sm$ has been opened by the X(5) model.
  Using the methods of Rowe {\em et al.} \cite{Rowe2004a, Rowe2004b, Rowe2005}, which permit exact solution of the Bohr Hamiltonian for wide classes of $u(\beta,\gamma)$, a large scope for future work exists.
  In particular, the energies and $B(E2)$ values for the $K^\pi=0_2^+$ band need careful exploration.
  Also, the $\beta$ and $\gamma$ dependence of the moments of inertia need to be explored.
   
\begin{acknowledgments}

  We wish to thank colleagues at the LBNL 88-Inch Cyclotron and OSU reactor for
assistance in the experiments and David Rowe and Paul Garrett for critical reading of this manuscript.  
  This work was supported in part by DOE grants/contracts
DE-FG02-96ER40958 (Ga Tech), DE-FG03-98ER41060 (OSU), DE-AC03-76SF00098
(LBNL), and DE-FG02-96ER40978 (LSU) and by the NSF REU program award PHY-0137451 (Ga Tech).

\end{acknowledgments}


\end{document}